\documentclass[nofootinbib, tightenlines, eqsecnum, floats, aps,amsfonts,twocolumn]{revtex4}

\usepackage{amsgen,amsbsy,amsmath}
\usepackage{mathrsfs}
\usepackage{graphicx}
\usepackage{xcolor}
\usepackage[hidelinks]{hyperref}
\usepackage[normalem]{ulem}
\usepackage{enumerate}

\newcommand {\be} {\begin{equation}}
\newcommand {\ee} {\end{equation}}

\newcommand{\dd}{\mbox{d}}
\newcommand{\tq}{\tilde{q}}

\newcommand{\tl}{\theta}

\usepackage{color}

\begin{document}

%\title{Multiply self-intersecting MOTS and extreme mass ratio mergers}
\title{MOTS in Schwarzschild: multiple self-intersections and extreme mass ratio mergers}
%\title{MOTS with multiple self-intersections and  extreme mass ratio mergers}
%\title{A menagerie of marginally outer trapped (open) surfaces in Schwarzschild spacetime}
%\title{Marginally outer trapped (open) surfaces and extreme mass ratio mergers}
\author {Ivan Booth\footnote{E-mail:
ibooth@mun.ca}}
\affiliation{
Department of Mathematics and Statistics, Memorial University of
Newfoundland \\ 
St. John's, Newfoundland and Labrador, A1C 5S7, Canada}
\affiliation{Institut de Math\'ematiques, Universit\'e de Bourgogne, Dijon, 21000, France}
\author{ Robie A. Hennigar\footnote{E-mail: rhennigar@mun.ca}}
\affiliation{
Department of Mathematics and Statistics, Memorial University of
Newfoundland \\ 
St. John's, Newfoundland and Labrador, A1C 5S7, Canada}
\author{Saikat Mondal\footnote{E-mail: saikatm@mun.ca}}
\affiliation{
Department of Physics and Physical Oceanography, Memorial University of
Newfoundland \\ 
St. John's, Newfoundland and Labrador, A1B 3X7, Canada}

\begin{abstract}
We study the open and closed axisymmetric marginally outer trapped surfaces contained  in leaves of constant Painlev{\'e}-Gullstrand time for Schwarzschild spacetimes. We identify a family of closed MOTS in the black hole interior characterized by an arbitrary number of self-intersections. This suggests that the self-intersecting behaviour reported in [Phys.~Rev.~D 100, 084044 (2019)] may be a far more generic phenomenon than expected. We also consider open surfaces, finding that their behaviour is highly constrained but includes surfaces with multiple self-intersections inside the horizon. We argue that the behaviour of open MOTS identifies and constrains the possible local behaviour of MOTS during extreme mass ratio mergers.
%We study the open and closed axisymmetric marginally outer trapped surfaces contained  in leaves of constant Painlev\'e-Gullstrand time for Schwarzschild spacetimes. 
%While there are an infinite number of open surfaces through each point in spacetime, their possible behaviours are found to be strongly constrained.
%Notably, those behaviours include surfaces that self-intersect an arbitrary number of times inside $r=2m$. Among such surfaces we identify an infinite
%set of closed marginally outer trapped surfaces (MOTS) that can be uniquely identified by their number of self-intersections. 
%We argue that these results identify and constrain possible local behaviours of MOTS during extreme mass ratio mergers. 
\end{abstract}

\maketitle

%\title{Open MOTS in Schwarzschild and EMR mergers}
%\author[Booth]{I.~Booth$^1$, R.~Hennigar$^1$, S.~Mondal$^2$ }
%\address{$^1$Dept.~of Mathematics and Statistics, Memorial University, NL, Canada}
%\address{$^2$ Dept.~of Physics and Physical Oceanography, Memorial University, NL, Canada}
%%\address[HK]{Department of Mathematics and Statistics, Memorial University of Newfoundland}
%\ead{ibooth@mun.ca, \dots}

\section{Introduction}
\label{Sec:introduction}

In four-dimensional spacetime, a marginally outer trapped surface (MOTS) is a two-dimensional closed spacelike surface with vanishing outward null expansion. 
The best known example is a two-dimensional slice of the $r=2m$ Schwarzschild horizon. However slices of horizons from other stationary spacetimes are also MOTS. 
These include all black hole horizons in the Kerr-Newman family (outer and inner) as well as (past) cosmological horizons. 

When first introducing MOTS to students, it is common to assign them the problem of showing that one or more of these standard horizons is indeed marginally outer trapped.\footnote{At least that is 
what one the authors of this paper always does!} Those first calculations are always coordinate-adapted: one  calculates the null expansions for surfaces of constant time and radial coordinate and then demonstrates that on the horizons
those expansions vanish. This shows that the horizons are the only coordinate adapted MOTS but of course it does not say anything about more general surfaces. 

The full picture %, as is well-known in numerical relativity \cite{}, 
is significantly more complicated. 
Given a parameterization of any two-surface, its expansion is determined by a 
second order differential operator. As such, vanishing expansion means that the (parameterization of the) surface satisfies a corresponding second order partial differential equation. 
Equivalently, given any point in spacetime and any tangent-plane to that point, the second order equation can be used to integrate those initial conditions to a surface of vanishing null expansion. 
This is similar (and in special cases even equivalent) to the minimal surface problem in Riemannian 
geometry. As for the marginally outer trapped surfaces, there are an infinite number of minimal surfaces through each point, one for each tangent plane.

Typically, MOTS are searched for in the leaves of some time foliation and so are automatically  spacelike. Then, what determines which (if any) of the surfaces of vanishing expansion is a 
MOTS is whether or not they close. Unlike the spacelike and vanishing outward null expansion conditions, which can be determined point-by-point, closure is inherently global. One needs to know 
the full surface in order to be able to classify it as open or closed. However that closure can depend crucially on far-away (and spacelike separated) features of the geometry. 
In particular, one can imagine two locally equivalent spacetimes for which the same partial surface may or may not be part of a MOTS depending on remote geometric features that either do or do not cause it to ultimately close. 

Hence, while all parts of the definition are important, it is not unreasonable to think that we can learn much about the local properties of MOTS by studying surfaces of vanishing null expansion without worrying about whether or not they close.
Morally, this strategy is similar to that employed when studying the mathematical properties of event horizons. 
Locally, event horizons are surface-forming congruences of null geodesics but globally, future boundary conditions are necessary to 
identify the particular set of curves that form an event horizon. However, many properties of event horizons can be understood from the Raychaudhuri equation, which applies to any 
congruence of null curves. Analogously, by 
studying the general behaviours of marginally outer trapped \emph{open} surfaces (MOTOS) we can hope to learn about possible behaviours of MOTS. 
 The early parts of this paper will be devoted to such an investigation for Schwarzschild spacetimes.

The physical problem that originally motivated this study gives further impetus for studying these MOTOS. Consider an extreme mass ratio black hole merger of non-rotating black holes with 
the small black hole falling into the large black hole along an (approximate) timelike geodesic of the large black hole spacetime. Then close 
to the small black hole, its gravitational field will be dominant. Hence to leading order, spacetime near the small black hole will be Schwarzschild. 
This will continue to be the case even as the small black holes crosses the large black hole horizon and remain so inside (until the final approach to the singularity). In the limit where the mass ratio
becomes infinite, that approximation becomes exact.

%dynamical spacetime that contains a non-rotating black hole. Further assume that the ``background'' gravitational fields are relatively weak close to the black hole  and 
%so spacetime is very nearly Schwarzschild there. This is a configuration often studied using perturbation theory and matched asymptotic expansions \cite{eric,others}. A particular 
%example is an extreme mass ratio black hole merger where, even as the small black hole crosses the massive black hole horizon, the gravitational field of the small black hole will be dominant 
%in some neighbourhood. 

%Emparan and Mart\'inez very successfully used this limit to study the evolution of event horizons during an extreme mass ratio merger \cite{Emparan:2016,Emparan:2017vyp} and the original motivation for 
%the current paper was to obtain the corresponding results  for MOTS/apparent horizons. In such a case, the MOTS associated with the large black hole 
%cannot be expected to close within the regime of Schwarzschild approximation and so we must necessarily study MOTOS. However even if these are understood perfectly,  there
%remains the complication of identifying
%which is the ``correct'' MOTOS that is the ``true'' geometric horizon. We will provide partial results towards this goal. 

{Emparan and Mart\'inez very successfully used this limit to study the evolution of event horizons during an extreme mass ratio merger \cite{Emparan:2016,Emparan:2017vyp} (see also 
\cite{emparan2020precursory} for an interesting extension to neutron star-black hole mergers). In their analysis, a congruence of null geodesics selected by its asymptotic properties  (e.g. that it be asymptotically planar) propagating in the Schwarzschild geometry  plays the role of the event horizon of the large black hole. Since solutions of the (null) geodesic equations in the Schwarzschild geometry can be obtained in terms of elliptic functions, these authors were able to obtain an ``exact description'' of the EMR merger. The original motivation for the current paper was to supplement the analysis of~\cite{Emparan:2016,Emparan:2017vyp} by understanding the properties of MOTS/apparent horizons.  In such a case, the MOTS associated with the large black hole 
cannot be expected to close within the regime of  the Schwarzschild approximation and so we must necessarily study MOTOS. However even if these are understood perfectly,  there
remains the complication of identifying
 the ``correct'' MOTOS that is the ``true'' geometric horizon. We will provide partial results towards this goal. 
}

{Our approach in this work will be to consider the possible behaviours of MOTOS within the Painlev{\'e}-Gullstrand slicing of the Schwarzschild spacetime. This slicing provides a number of 
advantages compared to the usual time-symmetric Schwarzschild slicing.  In Painlev\'e-Gullstrand, surfaces of constant time are spacelike and so any two-surface contained within such a slice will necessarily also 
be spacelike. However, in contrast to the time-symmetric slicing,  these coordinates are  horizon-penetrating which makes it possible to study  MOTOS that cross the horizon. Moreover, 
this slicing is non-static which has the consequence that the inward and outward expansions need not vanish simultaneously. These three features lift a degeneracy that is a special feature of
the usual Schwarzschild slicing and consequently we expect the results for the Painlev\'e-Gullstrand slicing to be more generic.  }
%{allow for a much richer set of possible behaviours for the MOTOS. }

%That said consider the case of an extreme mass ratio black hole merger. Then, very close to the small black hole we can expect the spacetime to be approximately Schwarzschild. Then MOTS 
%for the full spacetime should still be spacelike and marginally outer trapped in this regime but they will not close in the Schwarzschild regime. This is the original motivation for the study 
%of marginally outer trapped \emph{open} surfaces (MOTOS) that we present in this paper. The goal was to better understand the dynamics of MOTS during mergers by first understanding
%their possible local behaviours close to the a small black hole. Local dynamics can be studied without having access to the full MOTS. 

The paper is organized in the following way.  In Section \ref{Setup} we will set up the problem, deriving the formulae that define the axisymmetric MOTOS in Schwarzschild that are contained
in leaves of constant 
Painlev\'e-Gullstrand time.  While these equations can be solved exactly only  in the very simplest cases, Section \ref{analytic} applies 
analytic and perturbative techniques to learn as much as we can about the possible properties of those MOTOS. Section \ref{Sec:Numerics} solves the equations numerically 
and systematically works through possible solutions and their behaviours. This turns out to be an interesting problem in its own right and not as overwhelming as one might expect. A highlight of 
this section is the discovery of fully-fledged MOTS inside $r=2m$ that can have an arbitrary number of self-intersections. Section \ref{merger} returns to the original motivation and 
examines what the preceding sections can tell us about the 
behaviour of MOTS during extreme mass ratio mergers. Section \ref{discuss} summarizes the results and looks forward to future works. Appendix \ref{Sec:AB} is a technical section that examines sub-leading order asymptotic behaviours of MOTOS in Schwarzschild Painlev\'e-Gullstrand. 

A partial study of some of these MOTOS appeared in \cite{Booth:2017fob} in the context of a study of horizon stability. However we now realize that there were problems with the 
MOT(O)S generating algorithms  in that paper. Those algorithms were unable to track loops in the MOT(O)S and instead, when faced with one, incorrectly showed the surface diving into 
the singularity. Hence the exotic looping structures seen in this paper were missed in that earlier one.

\section{MOTOS in Schwarzschild: Painleve-Gullstrand coordinates}
\label{Setup}

%\subsection{Definitions}

\subsection{General considerations}
\label{general}

For the rest of this paper we restrict our attention to two-dimensional spacelike surfaces $S$ in the four-dimensional Schwarzschild
spacetime. Then, at any point, the normal space to such a surface is two-dimensional and  timelike and so can be spanned by two null vectors which we label $\ell^+$ and $\ell^-$. We take both 
vectors to be future 
pointing (there is no ambiguity of past and future for the cases we study) with $\ell^+$ being outward and $\ell^-$ being inward pointing 
(for some of the surfaces that we consider, this labelling is ambiguous but we will 
deal with such problems as they arise). For convenience we cross-scale so that 
\begin{equation}
\ell^+ \cdot \ell^- = -1 \, .
\end{equation}
The remaining one degree of scaling freedom, $\ell^+ \rightarrow e^f \ell^+$ and $\ell^- \rightarrow e^{-f} \ell^-$ for a free function $f$, will be 
irrelevant in this paper.

The full spacetime metric  $g_{\alpha \beta}$ will induce a spacelike two-metric $\tilde{q}_{AB}$ on $S$ that satisfies:
\begin{equation}
\tilde{q}^{AB} e^\alpha_A e^\beta_B = \tilde{q}^{\alpha \beta} = g^{\alpha \beta} + \ell^{+ \alpha} \ell^{- \beta} + \ell^{- \alpha} \ell^{+ \beta}
\end{equation}
and the expansions associated with these normals are
\begin{equation}
\theta^{+} = \tilde{q}^{\alpha \beta} \nabla_\alpha \ell^+_\beta \; \; \mbox{and} \; \;  \theta^{-} = \tilde{q}^{\alpha \beta} \nabla_\alpha \ell^-_\beta \; .
\end{equation}
In these expressions greek letters and capital latin letters are respectively spacetime and surface indices and $e_A^\alpha$ is the pull-back/push-forward operator between the spaces. 

We define a \emph{marginally outer trapped open surface} (MOTOS) to be an open spacelike surface with (at least) one normal direction of vanishing null expansion. We will refer to that 
direction as outward.\footnote{The ``outer'' in these names is not ideal. The nomenclature was developed  in the context of single black holes with non self-intersecting
horizons for which the notions of in and out were obvious. In the current context with multiple surfaces, some of which may have many self-intersections, this notion is much less clear. However
we keep this historical notation (partly because marginally trapped surface already has another common meaning). In this paper ``outer'' really just means that the expansion vanishes in at least one direction.}
If we are referring to a general case where the surface might be either open or closed we will use the somewhat awkward term MOT(O)S. 

Note that it is possible for $\theta_+$ and $\theta_-$ to  vanish simultaneously. For example, in standard Schwarzschild coordinates
\begin{align}
g_{\alpha \beta} \dd x^\alpha \dd x^\beta = &  - \left( 1 - \frac{2m}{r} \right) \mbox{d} t^2 +  \left( 1 - \frac{2m}{r} \right)^{-1} \mbox{d} r^2\\ 
&  + r^2 \left( \dd \theta^2 + \sin^2 \! \theta \dd \phi^2 \right) \, , \nonumber
\end{align}
hypersurfaces $\Sigma_t$ of constant $t$ with unit normal 
\be
\hat{u}_{\alpha} \dd x^\alpha = - \sqrt{1 - \frac{2m}{r}} \dd t
\ee
have vanishing extrinsic curvature: $K_{\alpha \beta} = 0$. Then a two-surface in $\Sigma_t$ with spacelike normal $ \hat{r} _\alpha$ will have null normals (up to the usual rescaling)
\be
\ell^+ = \hat{u} + \hat{r} \; \; \mbox{and} \; \; \ell^- = \frac{1}{2} \left( \hat{u} -  \hat{r}  \right) 
\ee
and so 
\begin{align}
\theta^+ & = \tilde{q}^{\alpha \beta}  \nabla_\alpha (\hat{u}_\beta +  \hat{r} _\beta) = \tilde{q}^{\alpha \beta} K_{\alpha \beta}  +  \tilde{q}^{\alpha \beta} \nabla_\alpha   \hat{r} _\beta \nonumber \\
& = \tilde{q}^{\alpha \beta} \nabla_\alpha   \hat{r} _\beta 
\end{align}
while 
\begin{align}
\theta^- & =  \frac{1}{2} \tilde{q}^{\alpha \beta}  \nabla_\alpha (\hat{u}_\beta -  \hat{r} _\beta) =  \frac{1}{2} \tilde{q}^{\alpha \beta} K_{\alpha \beta}  - \frac{1}{2}  \tilde{q}^{\alpha \beta} \nabla_\alpha   \hat{r}_\beta \nonumber \\
& = - \frac{1}{2} \tilde{q}^{\alpha \beta} \nabla_\alpha   \hat{r} _\beta \, . 
\end{align}
Hence if one of these vanishes, both vanish: MOT(O)S are minimal surfaces in the  $\Sigma_t$. Rescalings of the null vectors similarly scale the expansions and so do not affect whether or not 
they vanish. 

While this is an interesting situation in its own right (which we will return to in Section \ref{discuss}) it is also a very special case.  Most coordinate systems (and in particular those for dynamical 
spacetimes) do not have time slices of vanishing extrinsic curvature. Hence, in the upcoming sections we will  work in a coordinate system with non-stationary slices. Then, for each point
in space and each tangent plane there will actually be two MOT(O)S: one for $\ell^+$ and one for $\ell^-$.

Of course  there are many more MOT(O)S than those found in the leaves of any particular foliation. Hence we do not  claim that all MOT(O)S in a Schwarzschild spacetime behave in the
same way as the subset that we have studied. Testing the generality of our results will be left for a future paper.
%, however it should be noted that identifying MOTS in the leaves of a single foliation is
%the way that they are usually studied (for example in numerical relativity). 

\subsection{Schwarzschild Painlev\'e-Gullstrand}
\label{setup}

{ We choose to work in  Painlev\'e-Gullstrand coordinates so that  the time slices are: 1) spacelike (hence any two-surface in them will also be spacelike), 2) horizon penetrating
(so we can study MOT(O)S that cross $r=2m$) and 3) non-static (inward and outward null expansions will not vanish simultaneously). As noted in both the introduction and previous section, 
we believe that these
properties mean the behaviours of  the MOT(O)S found in this slicing will be representative of those found in a generic slicing (as opposed to that of standard, static, Schwarzschild coordinates).  }

As a  bonus, since the hypersurfaces of constant Painlev\'e-Gullstrand
time are intrinsically Euclidean $\mathbb{R}^3$ we can use 
Cartesian coordinates,
$x = r \sin \theta$ and $z = r \cos \theta$, 
 to describe curves in the $\phi=0$ plane and understand them in the usual Euclidean way. 
 
%
%
%, it is easy to interpret  the geometry of MOT(O)S depicted in diagrams. 
%We often use 
%This will even be simpler as we restrict our attention to axisymmetric MOT(O)S. 

\subsubsection{Metric and normals}

With greek letters running over $\{\tau,r,\theta,\phi\}$, the Schwarzschild metric in Painlev\'e-Gullstrand coordinates takes the form~\cite{Martel:2000rn}:
\begin{align}
g_{\alpha \beta} \dd x^\alpha \dd x^\beta  = & - \left( 1 - \frac{2m}{r} \right) \dd \tau^2 + 2 \sqrt{\frac{2m}{r}} \dd \tau \dd r \\
& +  \dd r^2 + r^2  \left( \dd \theta^2 + \sin^2 \! \theta \dd \phi^2 \right) \nonumber \, .
\end{align}
As noted above, the $\Sigma_\tau$ slices of constant $\tau$ are intrinsically flat with metric
\be
h_{ij} dx^i dx^j =   \dd r^2 + r^2\left( \dd \theta^2 + \sin^2 \! \theta \dd \phi^2 \right)   \, ,
\ee
and (non-flat) extrinsic curvature 
\be
K_{ij} \dd x^i \dd x^j =  \sqrt{\frac{m}{2r^3}}   \dd r^2 -    \sqrt{ 2mr}  \left( \dd \theta^2 + \sin^2 \! \theta \dd \phi^2 \right)  
\ee
which can be calculated from the future-oriented  unit timelike normal
\be
\hat{u} = \left( \frac{\partial}{\partial \tau}\right) -  \sqrt{\frac{2m}{r}} \left(\frac{\partial}{\partial r} \right) \, , 
\ee
to those slices (as a one-form this is $-\dd \tau$). In these expressions lower case latin letters run over $\{r,\theta,\phi\}$.
%
%
%
%
%which was calculated from the future-oriented  unit timelike normal to  $\Sigma_T$: 

An axisymmetric surface $S$ in a given $\Sigma_{\tau_o}$ can be parameterized by coordinates
$(\lambda, \phi)$ as
\be
(\tau_o,r,\theta,\phi) = (T_o,R(\lambda),\Theta(\lambda) , \phi) \,  ,
\ee
for some functions $R(\lambda)$ and $\Theta(\lambda)$. The tangent vectors to this surface are
\be
\frac{d}{d \lambda} = \dot{R} \left( \frac{\partial}{\partial r}  \right) + \dot{\Theta} \left( \frac{\partial}{\partial \theta} \right) \; \; \mbox{and} \;  \; \frac{\partial}{\partial \phi}
\, ,  \label{tangent}
\ee
with dots indicating derivatives with respect to $\lambda$. The induced two-metric is 
\be
\tilde{q}_{AB} \dd y^A \dd y^B =  (\dot{R}^2 + R^2 \dot{\Theta}^2) \dd \lambda^2 + (R^2 \sin^2 \! \Theta) \dd \phi^2 \, , 
\ee
with inverse:
\begin{align}
\tilde{q}^{AB} \left(  \frac{\partial}{\partial y^A}  \right) \otimes \left( \frac{\partial}{\partial y^B} \right)=&   \frac{1}{ \dot{R}^2 + R^2 \dot{\Theta}^2 }  \left( \frac{\partial}{\partial \lambda} \right)   \otimes \left( \frac{\partial}{\partial \lambda} \right) \nonumber \\
& + \frac{1}{R^2 \sin^2 \! \Theta} \left( \frac{\partial}{\partial \phi} \right) \otimes  \left( \frac{\partial}{\partial \phi} \right) \nonumber \, .  
\end{align}
Upper case latin letters run over $\{\lambda, \phi\}$.

From the tangent vectors (\ref{tangent}) it is straightforward to show that a unit spacelike normal to $S$ in $\Sigma_\tau$ is 
\be
\hat{r} = \frac{R}{\sqrt{\dot{R}^2 + R^2 \dot{\Theta}^2}} \left( \dot{\Theta} \left( \frac{\partial}{\partial r}  \right) 
 - \frac{\dot{R}}{R^2} \left( \frac{\partial}{\partial \theta}  \right)  \label{hr}
 \right) \; . 
\ee
Hence a suitable pair of null normals to $S$ is given by:
\be
\ell^+ = \hat{u} + \hat{r} \; \; \mbox{or} \; \; \ell^- =  \frac{1}{2} \left( \hat{u} - \hat{r} \right) \, .
\ee
As we will be looking for cases where the expansion associated with one of these vanishes, the remaining scaling freedom does not matter. 

The MOT(O)S that we deal with in the upcoming sections will often be quite complicated and necessarily 
covered by multiple coordinate patches. Then given that the orientation of  $\hat{r}$ in (\ref{hr}) has been tied to details of those parameterizations, it may (and will) be necessary 
to switch back and forth between $\ell^+$ and $\ell^-$ as we switch coordinate patches in order to maintain a consistent ``outward'' direction.

%Note that this points in the positive $r$ direction if $\dot{\Theta}>0$: that is if $\Theta$ increases as $\lambda$ increases. However the direction of vanishing null-
%expansion is not guaranteed to point in that direction
%
%
%In the following sections this will 
%not always be the case. Hence we may have to switch between working with $\pm \hat{r}$.

\subsubsection{Expansions}

The trace of the extrinsic curvature of $S$ in $\Sigma_\tau$ with respect to $\hat{r}$ is
\begin{align}
\theta_{(\hat{r})} & \equiv \tilde{q}^{ij} \nabla_i \hat{r}_j  \nonumber\\
&=  \tilde{q}^{\lambda \lambda} \left( \frac{\partial}{\partial \lambda} \right)^i \frac{\partial \hat{r}_i}{\partial \lambda} - \tq^{ij} \Gamma^k_{\phantom{k}ij} \hat{r}_k  \nonumber\\ 
&  =  \frac{1}{\sqrt{\dot{R}^2 + R^2 \dot{\Theta}^2}} \label{tr}  \\
 &\times \Bigg( \frac{R (\dot{R} \ddot{\Theta} - \ddot{R} \dot{\Theta})+\dot{\Theta} \dot{R}^2}{\dot{R}^2 + R^2 \dot{\Theta}^2} - \frac{\dot{R} \cot \Theta}{R} + 2 \dot{\Theta} \Bigg) \nonumber
\end{align}
where 
\be
\tilde{q}^{ij} = \left( \frac{\partial x^i}{\partial y^A} \right) \left(  \frac{\partial x^j}{\partial y^B}  \right) \tilde{q}^{AB}
\ee is the push-forward of the inverse surface metric into $\Sigma_\tau$. 
Note that as $S$ is a surface in Euclidean $\mathbb{R}^3$, no $m$ appears in this expression.

However the mass does show up in the extrinsic curvature of $S$ with respect to the unit timelike normal 
 $\hat{u}$. In that case 
 \begin{align}
\theta_{(\hat{u}) }   \equiv \tilde{q}^{\alpha \beta} \nabla_\alpha \hat{u}_\beta   & =  K_{\alpha \beta} h^{\alpha \beta} -  K_{\alpha \beta} \hat{r}^\alpha \hat{r}^\beta \nonumber  \\
& =  -\frac{\sqrt{2m}}{2 R^{3/2}} \left( \frac{\dot{R}^2 + 4 R^2 \dot{\Theta}^2}{\dot{R}^2 + R^2 \dot{\Theta}^2}\right) \; .  \label{tu}
\end{align}
Thus the two possible equations for vanishing null expansions are:
\begin{align}
\tl^+ = \theta_{(\hat{u})} + \theta_{(\hat{r})} = 0  \; \; \mbox{or} \; \; 
\tl^- = \frac{1}{2} \left( \theta_{(\hat{u})} - \theta_{(\hat{r})} \right) = 0  \;. 
\end{align}
That is 
\begin{align}
 \pm & \left(\dot{R} \ddot{\Theta} - \dot{\Theta}\ddot{R} + \frac{3 \dot{\Theta}{\dot{R}}^2}{R} - \frac{\dot{R} \cot\Theta}{R^2} \left(\dot{R}^2 + R^2 \dot{\Theta}^2 \right)  + 2 R \dot{\Theta}^3  \right) \nonumber \\
& - \sqrt{\frac{m}{2}}  \frac{\sqrt{\dot{R}^2 + R^2 \dot{\Theta}^2} \left(\dot{R}^2 + 4 R^2 \dot{\Theta}^2 \right) }{R^{5/2}} = 0 \label{tlpm}
\end{align}

Though we have left $\lambda$ general so far,  it is easiest to work with coordinate parameterizations. That is $\lambda = \theta$ or $\lambda = r$. Then we have four possible MOT(O)S generating equations. For functions $R(\theta)$ and $\Theta(r)$ and using subscripts to indicate derivatives they are:
\begin{align}
 R^{\mbox{\tiny{Eq}}}_\pm : \; \; & R_{\theta \theta}  - \frac{3R_\theta^2}{R} + \frac{R_\theta \cot \theta}{R^2} \left(R_\theta^2 + R^2  \right)   - 2R \label{REq} \\
&  \pm \sqrt{\frac{m}{2}}  \frac{ \sqrt{R_\theta^2 + R^2} \left(R_\theta^2 + 4 R^2 \right)  }{R^{5/2}} = 0  \nonumber 
\end{align}
for which 
\be
\hat{r} = \frac{R}{\sqrt{R_\theta^2+ R^2}} \left( \left( \frac{\partial}{\partial r}  \right) 
 - \frac{R_\theta}{R^2} \left( \frac{\partial}{\partial \theta}  \right)  
 \right) 
\ee
(that is points in the positive $r$ direction) and
\begin{align}
\Theta^{\mbox{\tiny{Eq}}}_\pm : \; \; \Theta_{rr} + \frac{3 \Theta_r}{r} - \frac{\cot \Theta}{r^2} \left( 1+ r^2  \Theta_r^2 \right) + 2 r \Theta_r^3 \label{ThetaEq}
\\
\mp \sqrt{\frac{m}{2}}  \frac{ \sqrt{1+ r^2 \Theta_r^2} \left(1 + 4 r^2 \Theta_r^2 \right)  }{r^{5/2}} = 0  \nonumber \end{align} 
for which
\be
\hat{r} = \frac{r}{\sqrt{1 + r^2 \Theta_r^2}} \left(\Theta_r \left( \frac{\partial}{\partial r}  \right) 
 - \frac{1}{r^2} \left( \frac{\partial}{\partial \theta}  \right) 
 \right)  
\ee
(that is points in the negative $\theta$ direction).  

As noted above, for a complicated MOT(O)S we may need to use all of these equations to generate an appropriate family of patches to fully cover it. The basic complication is
that it is not uncommon to have:
\begin{align}
\frac{\dd R}{\dd \theta} \rightarrow 0 \; \; &  \Leftrightarrow  \; \; \frac{\dd \Theta}{\dd r} \rightarrow \infty \; \; \mbox{or} \\
\frac{\dd \Theta}{\dd r} \rightarrow 0 \; \; &  \Leftrightarrow  \; \; \frac{\dd R}{\dd \theta} \rightarrow \infty  \; . 
\end{align}
The infinities are clearly  coordinate infinities:
% and do not imply  geometric discontinuities. Rather 
they are simply cases where the surface becomes tangent to either $\partial_r$ or 
$\partial_\theta$. Further, the fact that one derivative blows up as its reciprocal
goes to zero suggests a simple method for dealing with them: switch back and forth between solving for $R(\theta)$ and $\Theta(r)$ to always avoid singularities. \begin{figure}
\includegraphics[scale=.4]{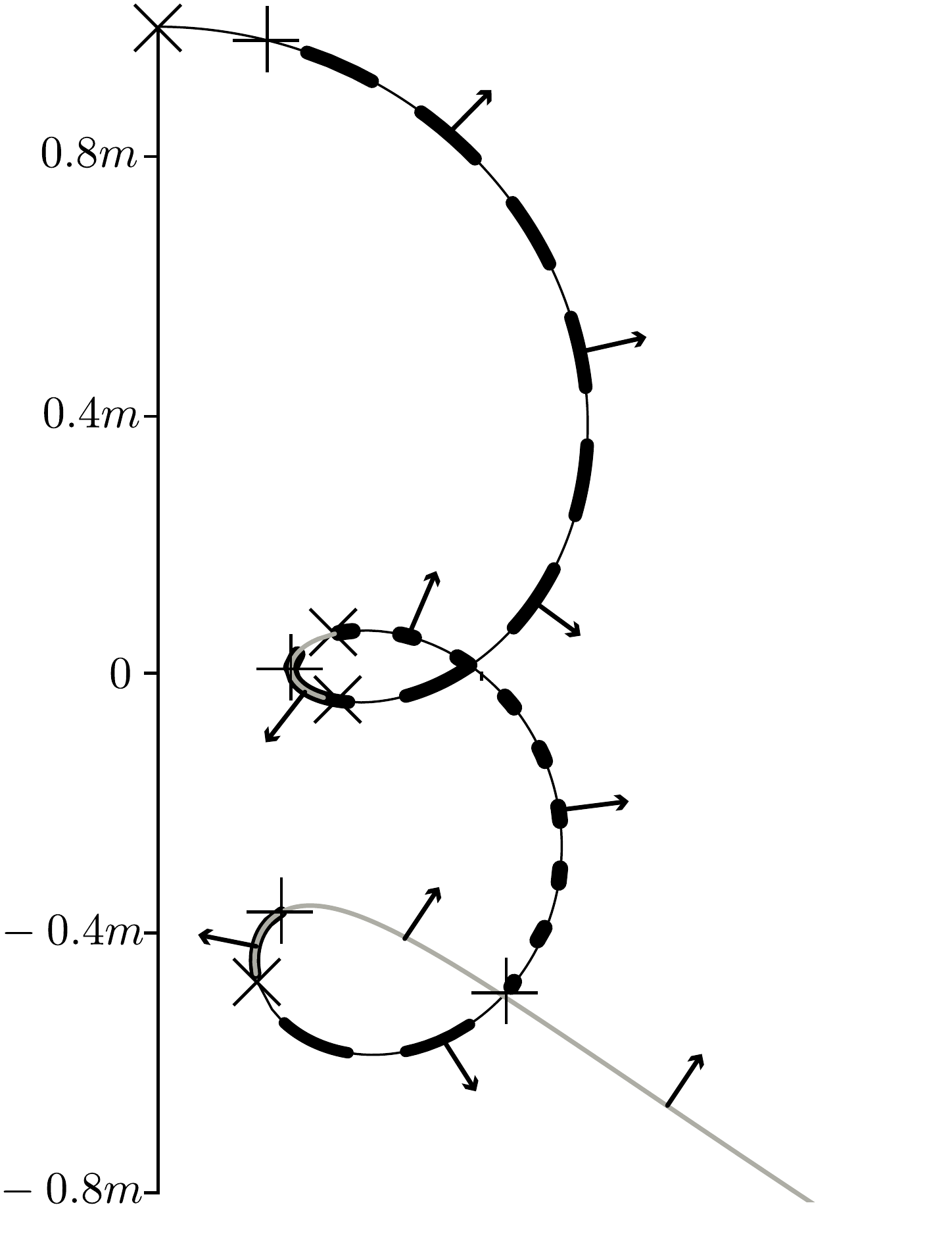}
\caption{Cross-section of a (doubly) self-intersecting MOTOS starting at $r=m$ and $\theta =0$ (the $z$-axis is shown but the $x$ is omitted for clarity). The direction of vanishing null expansion is shown by the arrows.
This surface is 
built out of seven patches. Starting from the $z$-axis these are solutions of: 1) $R_{\mbox{\tiny{Eq}}+}$ (thin black line), 2) $\Theta_{\mbox{\tiny{Eq}}-}$ (thick dashed line), 3) $R_{\mbox{\tiny{Eq}}-}$ (medium gray line), 4) $\Theta_{\mbox{\tiny{Eq}}+}$ (thick dotted line), 5) $R_{\mbox{\tiny{Eq}}+}$ (thin black line), 6) $\Theta_{\mbox{\tiny{Eq}}-}$ (thick dashed line) and 7) $R_{\mbox{\tiny{Eq}}-}$ (medium gray line). There is a substantial overlap of adjacent patches. $\times$s mark points where $\dd \Theta / \dd r \rightarrow \infty$  and $+$s are points where $\dd R / \dd \theta \rightarrow \infty$.  }
\label{Fig:Patches}
\end{figure}

This procedure is demonstrated in FIG.\ref{Fig:Patches} which shows a doubly self-intersecting MOTOS covered by seven patches. It starts from the 
$z$-axis at $(r,\theta) = (m,0)$ with a solution of $R^{\mbox{\tiny Eq}}_+$ and then cycles through $\Theta^{\mbox{\tiny Eq}}_-$, $R^{\mbox{\tiny Eq}}_-$ and $\Theta^{\mbox{\tiny Eq}}_+$ 
before starting again. As can be seen in the figure, the patches significantly overlap and individual equations only fail at the 
locations of coordinate discontinuities (marked with $\times$ or $+$). 

We will make repeated use of this procedure to generate MOT(O)S in Section \ref{Sec:Numerics}. First however we consider what we can learn analytically.

\section{Analytical results}
\label{analytic}

\subsection{Minkowski limit $m=0$ }
\begin{figure}
\begin{center}
\includegraphics[scale=0.5]{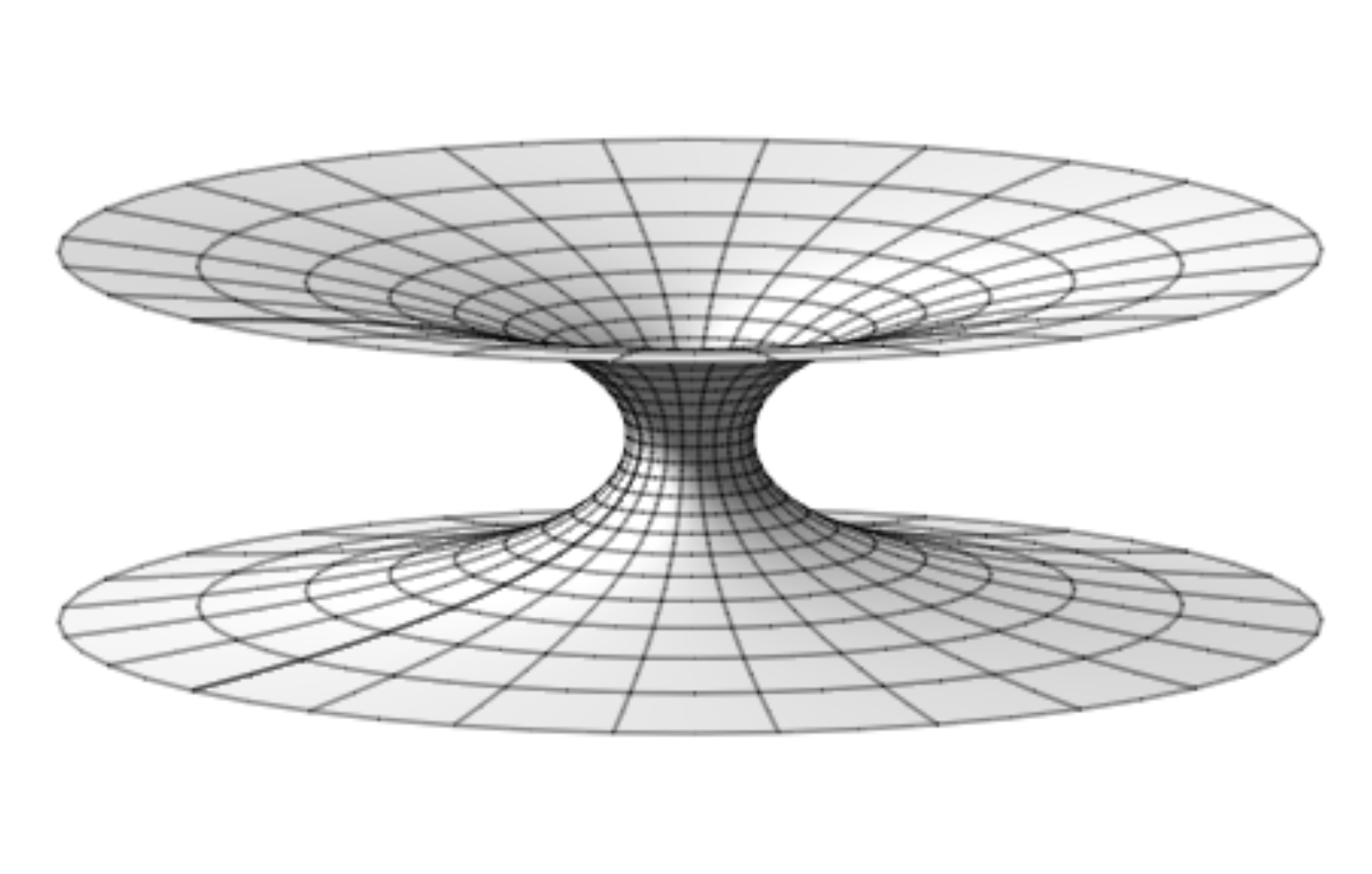}
\end{center}
\caption{A typical catenoid in Euclidean $\mathbb{R}^3$. The shape is characteristically hyperbolic with the sharpest principal curvatures around the narrow waist. }
\label{Cat3d}
\end{figure}

First, consider the simplest possible case. 
%We start with the simplest possible case in order to build some intuition about what kind of MOTOS we might expect. 
For $m=0$, $\theta_{(\hat{u})} = 0$ and so the problem reduces to solving $\theta_{(\hat{r})}=0$: that is finding rotationally symmetric minimal surfaces in 
Euclidean $\mathbb{R}^3$. 

These are well-known. Recall  that a surface is minimal if its two principal curvatures are 
equal in magnitude but opposite in orientation. The degenerate case is a $z=\mbox{constant}$ plane
for which both curvatures vanish, but the more interesting case is a catenoid for which the
 principal curvatures are non-zero. The principal curvature associated with the 
rotational symmetry is  oriented inwards towards the $z$-axis and so the other must 
be outwards. These opposite orientations give catenoids their characteristic hyperbolic shape that is 
shown in Figure~\ref{Cat3d}. The sharpest curvatures are 
necessarily around the waist of the catenoid (which has the smallest radius). 

\begin{figure}
\includegraphics[scale=0.8]{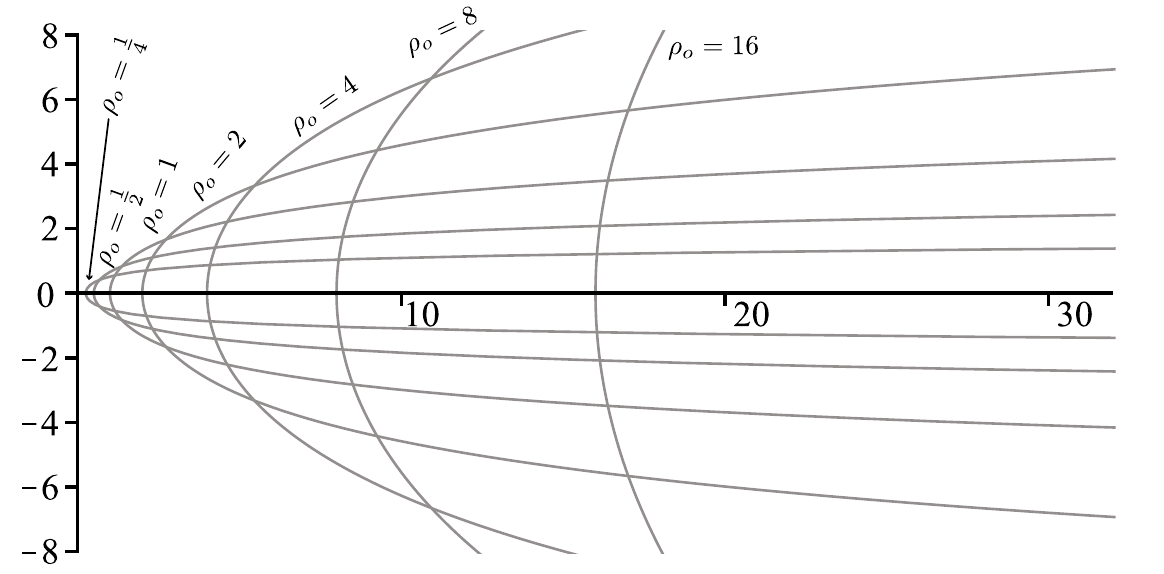}
\caption{Cross-sections of minimal surfaces with $z_o = 0$ for Euclidean $\mathbb{R}^3$. Note that as $\rho_o \rightarrow 0$ the catenoid reduces to the $z=0$ plane. For $z_o \neq 0$ the 
surfaces are appropriately shifted up or down in the $z$ direction.   }
\label{Fig:Catenoids}
\end{figure}

Quantitatively,  in cylindrical coordinates $(\rho, \phi, z)$ a catenoid  can be parameterized as
\be
\rho = \rho_o \cosh \left(\frac{z-z_o}{\rho_o} \right)  \; \; \mbox{for }  \; \; -\infty < z < \infty \; , \;  -\pi < \phi < \pi
\ee
where $(\rho_o, \phi, z_o)$ is the circle of closest approach to the $z$-axis. Sample catenoids with $z_o = 0$ are shown in FIG~\ref{Fig:Catenoids}.  Note the sharper curvature for those that approach closer to the $z$-axis. In particular, in the limit $\rho_o \rightarrow 0$ the catenoid degenerates to become the plane $z= z_o$.  

Converting to spherical coordinates but continuing to parameterize with  $\lambda = z$ we obtain:
\begin{eqnarray}
R(z) & = &\rho_o \sqrt{  \cosh^2 \left(\frac{z-z_o}{\rho_o} \right)+ \left( \frac{z}{\rho_o} \right) ^2 } \label{Rz}\\
\Theta(z) & = & \arctan \left(\frac{\rho_o}{z}  \cosh \left( \frac{z-z_o}{\rho_o} \right)  \right) \label{Tz}
\end{eqnarray}
which, by direct substitution, can be confirmed to be a solution to (\ref{tlpm}) (with $m=0$). 

\subsection{General behaviours}
\label{Gen}

While we will need numerics to study the full solutions of (\ref{REq}) and (\ref{ThetaEq}), we can still find constraints on possible behaviours of those solutions. 
Our equations for rotationally symmetric MOT(O)S are equivalently equations for curves in the $x>0$ half-plane.
%\footnote{Keep in mind that $x$ and $z$ have their regular Euclidean meaning since the $\Sigma_T$ are just Euclidean $\mathbb{R}^3$.} 
Hence understanding the surfaces is equivalent to understanding those 
curves and in particular we can consider their possible endpoints and turning points. That is what we do in this section, starting with possible endpoints. 

\subsubsection{Intersections with the $z$-axis}

First can a MOT(O)S intersect the $z$-axis of rotational symmetry? The obvious difficulty that such surfaces must overcome is that $\cot \theta$ blows up at $\theta =0$ and $\theta = \pi$ and so technically 
neither (\ref{REq}) nor (\ref{ThetaEq}) are defined there.  However it is easy to see that there are analytic curves that do intersect this axis and we find them by Taylor expanding $R(\theta)$ 
around $\theta=0$ or $\theta = \pi$, substituting into (\ref{REq}), and working order-by-order to solve for the series expansion. 

Then the blow-up is removed if and only if $R'(0)$ (or $R'(\pi)$) vanishes: \emph{any such curve must intersect the $z$-axis at a right angle}. 
Equivalently, intersections with the $z$-axis must be 
such that there are no conical singularities in $S$.

To second order with $\hat{r}$ pointing in the positive $r$ direction, 
we find
\begin{align}
R^{+}_0 (\theta)  &= R_o +  \frac{\sqrt{R_o} (\sqrt{R_o} - \sqrt{2m} ) }{2} \theta^2 + \mathcal{O} \left(\theta^4\right)  \label{nT1} \\
R^{-}_0 (\theta)  &=  R_o +  \frac{\sqrt{R_o} (\sqrt{R_o} + \sqrt{2m} ) }{2} \theta^2 + \mathcal{O}\left( \theta^4 \right) \; .  \label{nT2}
\end{align}
%while
%\begin{align}
%R^{+}_\pi (\theta)  &= R_o +  \frac{\sqrt{R_o} (\sqrt{R_o} - \sqrt{2m} ) }{2} (\pi- \theta)^2 + \mathcal{O} \left( (\pi-\theta^4) \right)  \label{nT1} \\
%R^{-}_\pi (\theta)  &=  R_o +  \frac{\sqrt{R_o} (\sqrt{R_o} + \sqrt{2m} ) }{2} (\pi -  \theta)^2 + \mathcal{O}\left( (\pi - \theta^4) \right) \; .  \label{nT2}
%\end{align}
The expansion coefficients are the same around $\theta = \pi$ and we see that intersections are allowed for any value of $R_o$ (except perhaps 
$R_o = 0$ which we consider in the next subsection). 

Focusing first on $R^+$ note that, as would be expected, $R^+ = 2m$ is a solution. However for $R_o>2m$, $R^+$ increases while for $R_o < 2m$ it 
decreases. The horizon is an unstable fixed point of this equation. By contrast, for all inward oriented normals, $R^-$ increases as the curve moves away from $\theta = 0$.

%
%
%The expansion coefficients are the same at $\theta = 0$ and $\theta=\pi$. Note in particular that the curves can intersect the $z$-axis for
%any $R_o$ (except maybe $R_o=0$ where there are extra difficulties that we will examine below). The $r=2m$ MOTS is a solution of $R^{+}_0$
%
%
%
% at second order they evolve away from $r=2m$.
%
%
%
% That is, for $R_o>2m$ they increase and $R_o<2m$ they decrease. Of course $r=2m$ 
%itself is a MOTS and, as expected for $R_o=2m$, the second order terms in $R^{+}$ also vanish --- in the case of $R^{-}$ the normal vector is oppositely oriented.
%
%\rah{Update increase/decrease comments}
%
%Terms up to any order can be easily calculated but these will suffice for our purposes.
%
%\rah{Reminder about consistent $z$ notation; comment about $r = 2m$ solving the equation.}

%Thinking back to the previous section, for $m=0$ the curves that intersect the $z$-axis are $z=z_o$ planes. 
%
\subsubsection{Intersections with $r=0$?}
Next we attempt to use the same technique to explore whether or not there are analytical curves that run into $r=0$. Restricted as we are to axisymmetry,  intrinsically $r=0$ is a point like any other 
on the $z$-axis. However extrinsically the surface has a  geometric singularity. 

There is no Taylor series solution of the equations around $r=0$ but if we expand as 
\be
\Theta(r) = \sum_{i=0}^{\infty} a_i r^{i/2}
\ee
 and substitute into 
(\ref{ThetaEq}), again trying to solve order-by-order, we find the following solution:
\begin{align} 
\Theta^\pm(r) =& \theta_0 \pm \sqrt{\frac{r}{2 m}} \left[-2 + \frac{59}{6} \frac{r}{m} - \frac{21003}{80} \frac{r^2}{m^2} \right.
\nonumber\\
&\left.+ \frac{6211375}{448} \frac{r^3}{m^3} + \cdots  \right] \, ,
\end{align}
where $\theta_0$ must be either $0$ or $\pi$. Thus, up to reflection symmetry, the series is completely determined with no free parameters appearing, suggesting that the curve intersecting the origin is unique. Moreover, the curve makes a cusp as it intersects the axis. However as the origin corresponds to a spacetime curvature singularity, this point around which we are attempting to expand is not part of the geometry. 

Given that we are trying to expand around a singular point, it is not too surprising that 
the series in brackets appears to have vanishing radius of convergence: an analysis of successive terms suggests that they grow without bound. 
That said, the existence of the series expansion suggests we may be  dealing with a differentiable but non-analytic function. Rational polynomial approximations to the function (e.g. Pad{\'e} approximants) exhibit much better behaviour.  Solving for the behaviour of $z$ near $r = 0$,  it is the same independent of whether $\pm$ is used:
\be 
z \approx r - \frac{r^2}{m} + \cdots \, , \quad x = \pm \sqrt{\frac{2 r}{m}} \left[-r + \frac{21}{4} \frac{r^2}{m} \cdots \right]
\ee
The leading-order behaviours of $z$ and $x$ here seem to match well with our numerical results as the curves close in on $r = 0$. However we have not definitively demonstrated the 
existence of this singularity-entering curve.

% However we immediately run into a problem. For $\Theta^{\mbox{\tiny{Eq}}}_+$ we require
%\begin{align}
%0 =&  - \sqrt{\frac{m}{2}} \frac{1}{r^{5/2}} - \frac{\cot{\Theta_o}}{r^2} + \frac{\Theta_1 (4 + \cot^2 \Theta_o)}{r} \nonumber \\ 
%& -  \sqrt{\frac{m}{2}} \frac{9\Theta_1^2}{2 r^{1/2}} + \mathcal{O} (r^0) \; . 
%\end{align}
%The only solutions are for $m=0$, $\Theta_o = \pi/2$ and $\Theta_1= 0 $. In fact continuing the expansion, all higher orders of derivatives must also vanish: the only curve running into $r=0$, is 
%$\Theta = \pi/2$ in the flat space limit. 
%
%{For $m>0$ there is no analytic MOT(O)S-generating curve running into $r=0$.} 
%
%\rah{I find the following solution for $\Theta(r)$ near $r=0$:
%\be 
%\Theta^\pm(r) = \pm \sqrt{\frac{r}{2 m}} \left[- 2 + \frac{59}{6} \frac{r}{m} + \mathcal{O}\left((r/m)^2 \right) \right]
%\ee
%Due to the square root factor, this curve is necessarily non-analytic at $r = 0$ --- its derivatives do not exist there. Thus it can be considered a solution for non-zero $r$ only. Nonetheless, the limiting behaviour is clear. We can then solve for the behaviour of $z$ near $r = 0$, and it is the same independent of whether $\pm$ is used:
%\be 
%z \approx r - \frac{r^2}{m} + \cdots
%\ee}

\subsubsection{Local extrema of $R(\theta)$}
\label{localExR}

From the previous section, there is at most a single curve that reaches
 $r=0$. Hence a generic curve $R(\theta)$ must have a minima. However for interesting solutions (like FIG.~\ref{Fig:Patches}) to exist we also require maxima. We now examine 
constraints on such extrema. 

For an extrema $R(\theta_o) = R_o$,  $R_\theta (\theta_o)= 0$ and so from (\ref{REq}) we must have
\be
R_{\theta \theta} (\theta_o) = 2 \sqrt{R_o} \left( \sqrt{R_o} \mp \sqrt{2m} \right)  \label{Rex}
\ee
at those points. In this expression the $+/-$ is for surfaces of vanishing inward/outward oriented null expansions. 

If $m=0$ it is clear that extrema can only be minima of $R$. No maxima are possible. This is consistent with the Euclidean
exact solutions where the catenoids have minimum values of $R$ but no maximum values. 

For $m>0$ we must distinguish between inward and outward oriented MOTOS. Surfaces with vanishing inward (towards $r=0$) expansions 
cannot contain local maxima of $r$.
%
%If $m > 0$ then there are more possibilities. For surfaces wi inward (that is towards $r=0$) oriented null expansions (the $+$ branch of (\ref{Rex})) minima are again the only 
%possibilities. No maxima are possible and so any curve segment with $R_\theta>0$ necessarily continues to increase. 
However, the situation is more interesting for surfaces with vanishing outward  (away from $r=0$) null expansions. Any $R_o > 2m$ is necessarily a 
local minima while any $R_o < 2m$ can only be a maxima. $R_o = 2m$ is a saddle point (a continuous set of points for which $R_\theta = 0$). 

Putting these together we note that there cannot be any local maxima in $r$ outside of $r=2m$ and hence there cannot be any closed axisymmetric
 MOTS that extend outside of that surface. This is consistent with the well-known result that marginally trapped surfaces necessarily 
 reside entirely within the causal black hole region \cite{Wald} as well as the more recent geometric results of \cite{Carrasco:2008aa} 
 (that MOTS cannot extend into an exterior region with a timelike Killing vector field). 

%In particular, in agreement with \cite{Carrasco:2008aa},  this means that there cannot be any closed axisymmetric MOTS that extend outside of $r=2m$.
 
The results are consistent with our  analysis of curves intersecting the $z$-axis. 
For $R_o \approx 2m$, near-horizon outward oriented MOTOS will always ``peel away'' from the horizon as one moves away from the extrema. 
However, they also allow for much more exotic behaviours like those seen in FIG.~\ref{Fig:Patches} (and will be seen in many later examples). 
Inside $r=2m$, curves can curl around,  switching their orientations, and so have both radial maxima and minima.

\subsubsection{Local extrema of $\Theta(r)$} 

Finally consider possible turning points of $\Theta(r)$. For $\Theta(r_o)=\Theta_o$ and $\Theta_r (r_o) = 0$, we have from (\ref{ThetaEq}) that:
\be
\Theta_{rr}(r_o) = \frac{1}{r_o^2} \left( \cot{\Theta_o} \pm \sqrt{\frac{m}{2r_o}}   \right)\, . 
\ee
For $m=0$ we recover Euclidean results. For $0 < \Theta_o < \pi/2$ and so $\cot (\Theta_o) > 0$, extrema can only be minima. For
$\pi/2 < \Theta_o < \pi$ and so $\cot (\Theta_o) < 0$, extrema can only be maxima. These are the catenoids turning back from the $z$-axis. 
 
For $m > 0$ things are again more interesting. For surfaces whose direction of vanishing null expansion is towards $\theta=0$ and $0 < \Theta < \pi/2$ (or $\theta=\pi$ for 
$\pi/2 < \Theta < \pi$) the results from flat space remain qualitatively the same. 
However for the opposite orientation there are maxima for $0 < \Theta < \pi/2$ (and minima for $\pi/2 < \Theta < \pi$) if $|\cot \Theta_o| < \sqrt{{m}/{2r_o}}$. 

This is similar to the situations that we found for $R$ in the last subsection and so for a multi-patch covered surface we can have both maxima and minima of $\Theta$. 

%
%This makes possible the intricate curls of FIG.~\ref{Fig:Patches} as well as the examples of later sections. 

\subsection{The large-$r$ limit}

\subsubsection{Minkowski limit $m=0$}
Even for minimal surfaces in  Euclidean space there is no way to explicitly invert either (\ref{Rz}) or (\ref{Tz}) and so obtain
an explicit solution $R(\theta)$ or $\Theta(r)$ in terms of elementary functions. Here we consider the 
large-$r$ limit. 

For large $r$ we can perturbatively invert (\ref{Rz}) to obtain
%\begin{eqnarray}
%z & = & \rho_o \left( \frac{z_o}{\rho_o} +   \ln \left[ \frac{2r}{\rho_o} \right]  \right)   - \rho_o \left( \frac{1}{4} + \frac{1}{2} \left( \frac{z_o}{\rho_o} +   \ln \left[ \frac{2r}{\rho_o} \right]  \right)^2  \right) \left( \frac{\rho_o}{r} \right)^2
%\\ & & + \mathcal{O} \left( (\rho_o/r)^3 \right)  \nonumber
%\end{eqnarray}
\begin{align}
z_{\mbox{\tiny{flat}}}  = & \mp \rho_o \Bigg(   X - \frac{2X^2+1}{4} \left( \frac{\rho_o}{r} \right)^2
%& \phantom{\mp \Bigg( }  - \frac{8X^4-16X^3+8X^2-8X+3}{32} \left( \frac{\rho_o}{r} \right)^4 \Bigg) \nonumber \\
  + \mathcal{O} \left( \frac{\rho_o^4}{r^4} \right)   \Bigg)  \label{zex} 
\end{align}
 where 
\be
X = \frac{z_o}{\rho_o} +   \ln \left( \frac{2r}{\rho_o} \right) 
\ee
and the $\mp$ depends on whether we are considering the upper or lower branch of the catenoid ($-$ upper, $+$ lower).  Higher order terms can include powers of $X$ (and so
$\ln r$) in addition to powers of $1/r$.
Note that $\lim_{r\rightarrow \infty} X = \infty$ and so $z$ similarly diverges, though only logarithmically. 

This slow growth can be also be seen in the large-$r$ behaviour of 
(\ref{Tz}). Asymptotically expanding with (\ref{zex}) we obtain:
%\begin{eqnarray}
%\Theta (r) =  \frac{\pi}{2} - \left( \frac{z_o}{\rho_o} +   \ln \left[ \frac{2r}{\rho_o}\right] \right) \left(\frac{\rho_o}{r} \right) + \mathcal{O}  \left( (\rho_o/r)^3 \right)  \, , 
%\end{eqnarray} 
\begin{align}
\Theta_{\tiny \mbox{flat}\pm} (r) =   \frac{\pi}{2} &  \mp \left(  X \left( \frac{\rho_o}{r} \right) + \frac{2X^3-6X^2-3}{12} \left( \frac{\rho_o}{r} \right)^3  \right) \nonumber \\
&  + \mathcal{O}  \left( \frac{\rho_o^5}{r^5}\right)  \, , \label{TFlat}
\end{align} 
and so in the large-$r$ limit $\Theta \rightarrow \pi/2$.  By direct substitution it can be confirmed that this is a solution for $\Theta_{\mbox{\tiny{eq}}\pm}$ (\ref{ThetaEq}) at large $r$ with $m=0$.

\subsubsection{General case}
\label{LargeR}

%\rah{I seem to find that the first $\mp$ that occurs should really be a $\pm$ -- someone should confirm to shut me down. This then changes the $\pm$ to a $\mp$ in (3.15) below as well. (I am probably making a stupid error but I cannot find it...)}

Next consider $m > 0$. There we might expect the minimal surfaces far from the black hole to behave similarly to the $m=0$ case: an asymptotic series
involving powers of $r$ and $\ln(r)$. This initial ansatz is wrong; no such solution exists. However if we add in half-powers of $r$ we do obtain a solution. To order $1/r^2$ we find asymptotic solutions to
$\Theta^{\mbox{\tiny{Eq}}}_\pm$ (\ref{ThetaEq}) of the form:
%\begin{align}
%\Theta_{\mbox{\tiny{asympt}}}^\pm   \!= & \frac{\pi}{2}   \mp 2 \sqrt{2} \left( \frac{m}{r} \right)^{1/2} +  \tilde{X} \left(\frac{\beta}{r}\right)  \pm  \frac{10 \sqrt{2}}{3} \left( \frac{m}{r} \right)^{3/2}  \nonumber %+ \frac{2\tilde{X}^3-6\tilde{X}^2-3}{12} \left( \frac{\beta}{r} \right)^3   \nonumber\\ 
%\\
%&  \nonumber\\ 
%%&+ \left(  \frac{371\sqrt{2} }{15}  m^{5/2} - \sqrt{2} \left( \tilde{X}^2 - 2 \tilde{X} - \frac{7}{9} \right) m^{1/2} \beta^2 \right) \frac{1}{r^{5/2}} \nonumber \\
%%& - \frac{115 + 82 Z}{4} \left( \frac{m\beta^2}{r^3} \right) 
%& + (3\tilde{X} - 7) \left( \frac{\beta m}{r^2} \right) + \mathcal{O} \left( \frac{m^i \beta^j }{r^{5/2}} \right) 
%\end{align}
\begin{align}
\Theta_{\mbox{\tiny{asympt}}}^\pm   \!= & \frac{\pi}{2}   \pm 2 \sqrt{2} \left( \frac{m}{r} \right)^{1/2} +  \tilde{X} \left(\frac{\beta}{r}\right)  \mp  \frac{10 \sqrt{2}}{3} \left( \frac{m}{r} \right)^{3/2}  \nonumber %+ \frac{2\tilde{X}^3-6\tilde{X}^2-3}{12} \left( \frac{\beta}{r} \right)^3   \nonumber\\ 
\\
&  \nonumber\\ 
%&+ \left(  \frac{371\sqrt{2} }{15}  m^{5/2} - \sqrt{2} \left( \tilde{X}^2 - 2 \tilde{X} - \frac{7}{9} \right) m^{1/2} \beta^2 \right) \frac{1}{r^{5/2}} \nonumber \\
%& - \frac{115 + 82 Z}{4} \left( \frac{m\beta^2}{r^3} \right) 
& + (3\tilde{X} - 7) \left( \frac{\beta m}{r^2} \right) + \mathcal{O} \left( \frac{m^i \beta^j }{r^{5/2}} \right) 
\end{align}
where $i+j = 5/2$, one should consistently choose either the top or the bottom sign to get a solution of the $\Theta^{\mbox{\tiny{Eq}}}_\pm$ and 
\be
\beta \tilde{X} =  \alpha + \beta \ln r \, ,  \label{beta}
\ee
with $\alpha$ and $\beta$ as free constants which  distinguish the individual solutions. 

The non-$m$ terms match up with the vacuum case (\ref{TFlat}) 
but intriguingly the post-$\frac{\pi}{2}$ leading order term behaviour changes: there is now a dominant $r^{-1/2}$ term.
In particular this means that at large $r$:
\be
z^\pm_{\mbox{\tiny{asympt}}} = r \cos \Theta^\pm_{\mbox{\tiny{asympt}}} \approx \pm 2\sqrt{2m} \, r^{1/2} - \beta \tilde{X} + \mathcal{O} (r^{-1/2}) \, . \label{AsymptZ}
\ee
That is, even asymptotically the large-$r$ behaviour of the MOTOS is different from the $m=0$ case: $z$ diverges as $\sqrt{r}$ rather than $\ln r$. Note however that this means that the leading
order behaviour is universal and independent of the particular solution. 

Thus there is a qualitative difference between $m=0$ (Minkowksi space) and any $m > 0$. Even arbitrarily far from the origin any non-zero mass still has an effect. 

%
%Essentially this arises because for any $m > 0$, the $t$=constant surfaces are boosted relative to 
%Minkowski space.

Finally consider the case of a MOTOS that does not intersect the $z$-axis and so has two asymptotic ends. Then both ends must behave as (\ref{AsymptZ}) at large $r$. Consider traversing the 
generating curve. If the direction of vanishing null expansion is consistently oriented along the curve, one end will necessarily asymptote as $z^+_{\mbox{\tiny{asympt}}}$ while the other will 
asymptote as $z^-_{\mbox{\tiny{asympt}}}$. That is asymptotically it will behave as
\be
r_{\mbox{\tiny{asympt}}} \approx \frac{z_{\mbox{\tiny{asympt}}}^2}{8m} \, . \label{parabola}
\ee
We will see this behaviour in Section \ref{nonZ}. 

The only way to avoid these parabolic asymptotics would be for the direction of vanishing null expansion to switch somewhere along the curve. In that case at some point on the curve we would 
have to have $\theta_{(\ell)}^+ = \theta_{(\ell)}^- = 0$ or equivalently $\theta_{\hat{u}} = \theta_{\hat{r}} = 0$. From (\ref{tr}) and (\ref{tu}) this cannot happen smoothly as it would require
$\dot{R} = \dot{\Theta} = 0$. % {\bf (think about this a bit more)} 

\subsection{Near-MOTS limit}
\label{NearMOTS}

We can also consider the behaviour of curves close to $r=2m$, looking for perturbative solutions to (\ref{REq}) of the form:
\begin{equation}
R(\theta) = 2m \left( 1 + \rho(\theta) \right)
\end{equation}
with $| \rho | \ll 1$. To first order the equations become:
\begin{align}
R_{\mbox{\tiny{eq}}}^{+}: & \rho_{\theta \theta} + \cot \theta \rho_\theta - \rho = 0  \label{rhop}  \\
R_{\mbox{\tiny{eq}}}^{-}: & \rho_{\theta \theta} + \cot \theta \rho_\theta - 3\rho = 4 \label{rhom}
\end{align}  
which respectively define surfaces with the same or opposite vanishing null orientation to $r=2m$. These have solutions in terms of Legendre functions
\begin{align}
\rho^+ & = A^+ P_{l^+} (\cos \theta)  + B^+ Q_{l^+} (\cos \theta) \\
\rho^- & = - \frac{4}{3} +  A^- P_{l^-} (\cos \theta)  + B^- Q_{l^-} (\cos \theta)
\end{align}
where $A^{\pm}$ and $B^{\pm}$ are free constants and 
\be
l^+ = -\frac{1 + i \sqrt{3}}{2} \; \; \mbox{and} \; \; l^- = -\frac{1 + i \sqrt{11}}{2} \; . 
\ee

We are mainly interested in the case where $\rho \rightarrow \rho_o$ and $\rho_\theta \rightarrow 0$ as $\theta \rightarrow 0$ (that is where the curve intersects the positive $z$-axis). Then
$B^\pm = 0$ (the associated Legendre functions diverge for $\theta = 0$) and we have:
\begin{align}
\rho^+ & = \rho_o P_{l^+} (\cos \theta) \\
\rho^- & = - \frac{4}{3} +  \left(\frac{4}{3} + \rho_o \right) P_{l^-} (\cos \theta)  \; . 
\end{align}
It is immediately obvious that for $\rho_o = 0$ (that is $z_o = 2m$), $\rho^+ = 0$. This is in accord with the uniqueness theorem for MOT(O)S  \cite{Andersson:2009aa,Mosta:2015sga}: in this case two MOT(O)S 
touch with the same orientation of their directions of vanishing null expansion and so must be identical. Note too that for 
non-zero $\rho_o$ it will diverge by $\theta = \pi$, but that rate of divergence is controlled by the initial proximity to $r=2m$. 

By contrast $\rho^-$ diverges from $r=2m$,  even if  $\rho_o = 0$. For $\rho_o \ll 1$ that rate of divergence is essentially independent of $\rho_o$. Further 
$P_{l^-} (\cos \theta) > P_{l^+} (\cos \theta)$ for all $\theta$ and so for any $\rho_o$, $\rho^-$ diverges faster than $\rho^+$. These results will make more sense when seen in the 
context of the upcoming examples.

\section{Numerical Solutions}
\label{Sec:Numerics}

In this section we consider more general MOT(O)S and find a rich family of possible surfaces.  In Section \ref{FromBelow} and \ref{FromAbove} we will examine the possible rotationally symmetric
MOT(O)S that intersect the $z$-axis and which, at that point, have their direction of vanishing null expansion oriented in the positive $z$ direction. 
%In the third section we will examine the asymptotic behaviour of these surfaces.  
Due to the reflectional symmetry of Schwarzschild across
$z=0$, this is sufficient to understand all possible surfaces that intersect the $z$-axis. Section \ref{nonZ} will survey some of the MOTOS that do not intersect the $z$-axis. 

%A partial study of some of these surfaces appeared in \cite{Booth:2017fob} in the context of a study of horizon stability. However we now realize that there were problems with the 
%MOT(O)S generating algorithms  in that paper. Hence the exotic looping structures seen in the following sectsion were missed (and instead the MOT(O)S inside $r=2m$ were incorrectly believed 
%to end in the singularity). 

For surfaces departing from the $z$-axis, the coordinate representations break down: $\cot \theta$ diverges for $\theta=0$.
Hence we make use of the perturbative expressions (\ref{nT1}) or (\ref{nT2}) for an $R(\theta)$ leaving 
the $z$-axis to find $R$ and $R_\theta$ at some  small $\theta_o$. Then we have initial conditions  in a location where everything is well-defined. 

%\begin{align}
%& R^{z+}_0 (\theta)  = R_o +  \frac{\sqrt{R_o} (\sqrt{R_o} - \sqrt{2m} ) }{2} \theta^2 + \mathcal{O} \left(\theta^4\right)  \nonumber \\
%& \mbox{with} \; \; \frac{\dd}{\dd \theta} R^{z+}_0   = \sqrt{R_o} (\sqrt{R_o} - \sqrt{2m} ) \theta +  \mathcal{O} \left(\theta^3\right)
%\end{align}
%and
%\begin{align}
%& R^{z+}_\pi (\theta)  =  R_o +  \frac{\sqrt{R_o} (\sqrt{R_o} + \sqrt{2m} ) }{2} (\pi - \theta)^2 + \mathcal{O}\left( (\pi - \theta)^4 \right) \nonumber \\
%& \mbox{with} \; \; \frac{\dd}{\dd \theta} R^{z+}_\pi  = -\sqrt{R_o} (\sqrt{R_o} - \sqrt{2m} ) (\pi -\theta) +  \mathcal{O} \left(\theta^3\right)
%\end{align}
%respectively. These expressions can be used to obtain initial conditions a short distance from the axis where everything is well defined. 

\subsection{From below}
\label{FromBelow}

\begin{figure}
\includegraphics{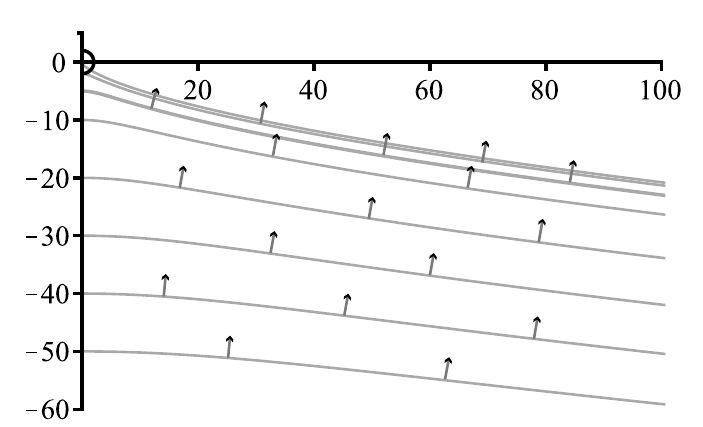}
\caption{Upward oriented rotationally symmetric MOTOS approaching (and passing) the $r=2m$ MOTS from below. The orientation of the null vector vanishing null expansion is indicated by the arrows. }
\label{BelowLarge}
\end{figure}

\begin{figure}
\includegraphics{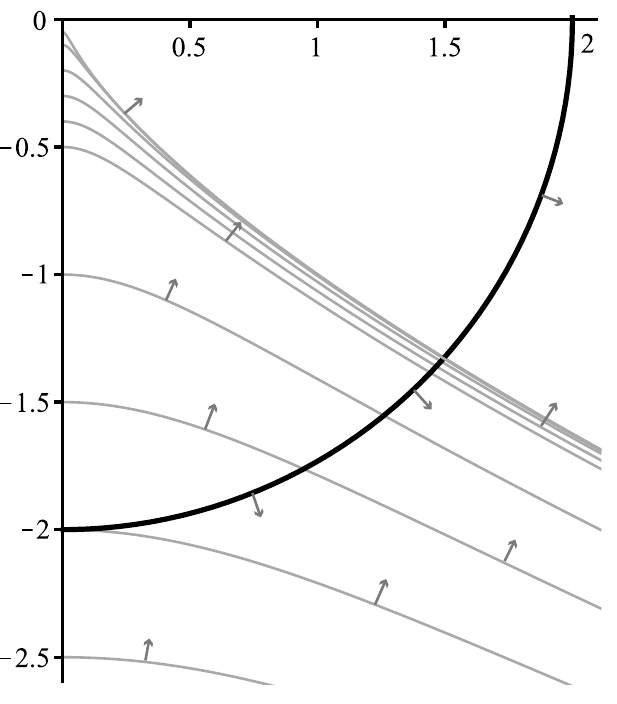}
\caption{Upward oriented rotationally symmetric MOTOS inside $r=2m$. The orientation of the null vector vanishing null expansion is indicated by the arrows. The MOTS and MOTOS can
be tangent to each other are $r=2m$ because they are oriented in opposite directions. %{\bf Is there something useful to say about the limiting curve?}
 }
\label{BelowSmall}
\end{figure}

We begin with the  case for which the MOT(O)S intersects the $z$-axis at $z_o<0$  with the direction of vanishing null expansion in the $+z$ direction. 
From our considerations in Section \ref{Gen} we expect there to be only minimum values of $R(\theta)$ for these surfaces and indeed this is what we find. 
Representative surfaces are depicted in FIG.~\ref{BelowLarge} and \ref{BelowSmall}. 
%\begin{figure}
%\includegraphics{FromBelowAsympt}
%\caption{Asymptotic behaviour of upward oriented rotationally symmetric MOTOS that intersect the $z$-axis with $z<0$. The vertical axis is the $z^+_{\mbox{\tiny{asympt}}}$ coordinate with 
%the universal $-2\sqrt{2mR}$ asymptotics removed. The starting location for the surfaces ranges from $r_o =0.1m$ to $r_o = 48.1m$ (hence their 
%staggered starting points). As expected their asymptotic behaviour is log-linear.  }
%\label{FBA}
%\end{figure}
%\begin{figure}
%\includegraphics{StraightLines}
%\caption{dd}
%\label{Bel}
%\end{figure}

This family of curves is quite simple but there are a couple of features to note. 
First the MOTOS can intersect  $r=2m$ and nothing particularly special happens at the point of intersection. The $z_o = 2m$ MOTOS is 
tangent to $r=2m$. As noted in the discussion of section  \ref{NearMOTS} this does not violate the
 uniqueness theorem for MOTOS: the horizon and $z_o = 2m$ MOTOS have opposite orientations at their point of contact. 
 
 The MOTOS are well behaved, maintaining their original ordering as they extend outwards from the $z$-axis to large $r$. That this ordering continues to be maintained asymptotically is confirmed in  Appendix~\ref{Sec:AB}. 
% We will return to this point in Section~\ref{Sec:AB}
% where we examine their asymptotic behaviour more carefully. Note too that in the approach to $r=0$ the MOTOS appear to asymptote together. 
% This asymptotic behaviour in more detail in Appendix~\ref{Sec:AB}.

\subsection{From above}
\label{FromAbove}

\subsubsection{Results}
\begin{figure}
\includegraphics[scale=.9]{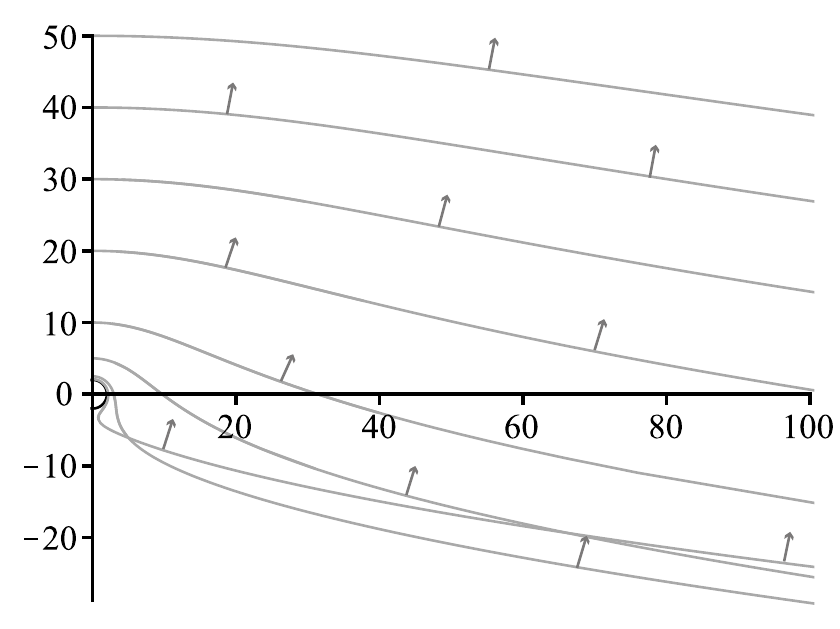}
\caption{Upward oriented rotationally symmetric MOTOS approaching (and passing) the $r=2m$ MOTS from above. The orientation of the null vector vanishing null expansion is indicated by the arrows. }
\label{FAL}
\end{figure}

\begin{figure}
\includegraphics{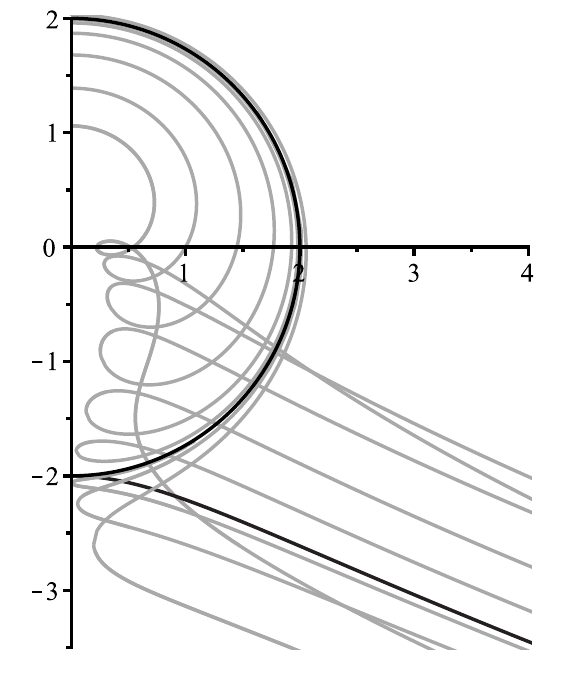}
\caption{Initially upward oriented rotationally symmetric MOTOS approaching (and passing) the $r=2m$ MOTS from above. Orientation vectors for the direction of vanishing null expansion are henceforth 
omitted but the direction can be tracked by following curves from the positive $z$-axis. The orientations of all curves is in the positive $z$ direction at both the $z$-axis and as they head off to infinity.  }
\label{Evolve1}
\end{figure}

The MOTOS originating from $z_o>0$ have a much more interesting set of behaviours than those with $z_o<0$.  For this case the tangent surface  at $z_o = 2m$ has the same orientation as $r=2m$ and 
so by the uniqueness theorem is identical. 
 As can be seen in FIG.~\ref{FAL} and \ref{Evolve1},  this uniqueness is the endpoint of a continuous process: MOTOS with $z_o \rightarrow 2m$ wrap  more and more closely to $r=2m$ during
 the approach (as would be expected from the results of Section \ref{NearMOTS}). They then  have to make sharper and sharper turns  to avoid $\theta = \pi$. The limit as $z_o \rightarrow 2m$ is the MOTS at $r=2m$ along with the oppositely oriented $z_o=2m$ 
 MOTOS (the black curves in FIG.~\ref{Evolve1}). From the perspective of the generating curves the limit is continuous. However the limiting curve itself is not smooth. 

The curves on either side of $z_o = 2m$ make their turns to avoid the $z$-axis in different ways. For $z_o>2m$ they turn counter-clockwise (to their left if one is moving along the curve starting from $\theta=0$). 
However for $z_o<2m$, the turn is clockwise (to their right) and they then self-intersect before exiting through $r=2m$. In both cases  they end up moving off to large $r$ with their 
direction of vanishing null expansion oriented in the positive $z$ direction. % (and so will asymptote to $z = - 2 \sqrt{2mr}$). 

\begin{figure}
\includegraphics{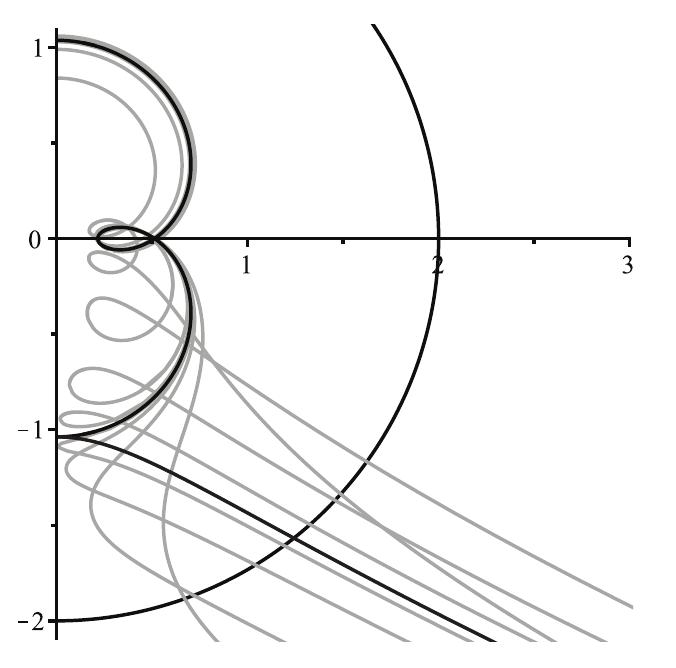}
\caption{Initially upward oriented rotationally symmetric MOTOS on either side of the one-loop MOTS.  Note that the second loop develops in a way that is qualitatively similar to the development of the first. The orientations of all curves is in the positive $z$ direction at both the $z$-axis and as they head off to infinity. }
\label{Evolve2}
\end{figure}

%
%The evolution of the MOTOS as $r_o$ decreases from being larger than to $2m$ to being smaller is shown in the left-hand side of FIG.~\ref{Evolve}. 
%Note how the wrapping seen  in FIG.~\ref{FAL} becomes tighter as $r_o \rightarrow 2m$ from above with a sharper and sharper turn from the $z$-axis. In the limit we have $r=2m$ plus 
%the upwards pointing surface $r_o=2m$ surface from the last section (the black curves).
%
MOTOS geometries become more complicated as $z_o$ further decreases. As can be seen in FIG.~\ref{Evolve1}, the loop grows, pulls back from $r=2m$ and migrates towards $z=0$. 
%At the same time, the part of the curve heading out to infinity begins to pull back towards the $z$-axis (as we saw for the outgoing branch during the approach to $2m$). 
This continues in FIG.~\ref{Evolve2} where the free end pulls back towards the $z$-axis with the turn becoming sharper and sharper until 
%The turn from the $z$-axis  becoming sharper and sharper until, as seen in FIG.~\ref{Evolve2}, 
for $z_o  = z_1 \approx 1.037m$ the curve intersects the $z$-axis at a right angle: this limit is a closed surface and so a (self-intersecting)  MOTS. It pairs with the  upwards pointing surface from the last section that  intersects at $z_o= -z_1$ to form the limit surface. 

\begin{figure*}
\includegraphics{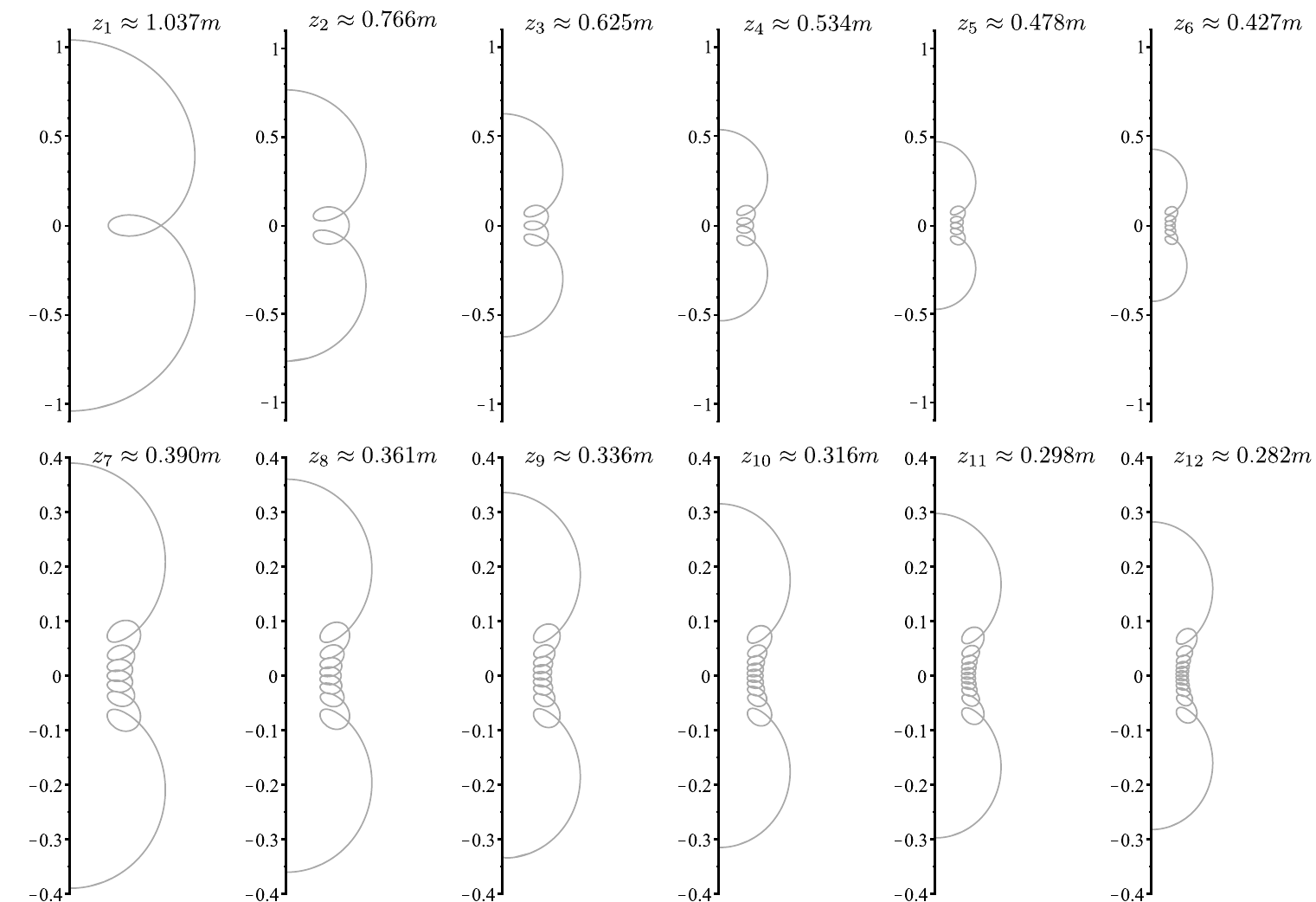}
\caption{The first twelve (post-$r=2m$) rotationally symmetric MOTS living in $t=\mbox{constant}$ slices of Schwarzschild in Painelev\'{e}-Gullstand coordinates.   
Axes labels are in units of $m$ and the subscript refers to the number of loops.  The $z_i$ values were obtained from numerical experiments: essentially using the shooting method for solving ODEs. 
There appear to be an infinite number of these surfaces. Note that the scale changes between the first and second rows.}
\label{FIG_MOTS12}
\end{figure*}

These general patterns repeat as we move deeper towards the singularity.  For $z_o < z_1$ a second loop forms which then also migrates towards $z=0$ (FIG.~\ref{Evolve2}). 
Ultimately for $z_o =z_2 \approx 0.766$ the two loops end up symmetrically spaced around $z=0$ and we have another MOTS (the second curve in FIG.~\ref{FIG_MOTS12}).  

\begin{figure}
\includegraphics{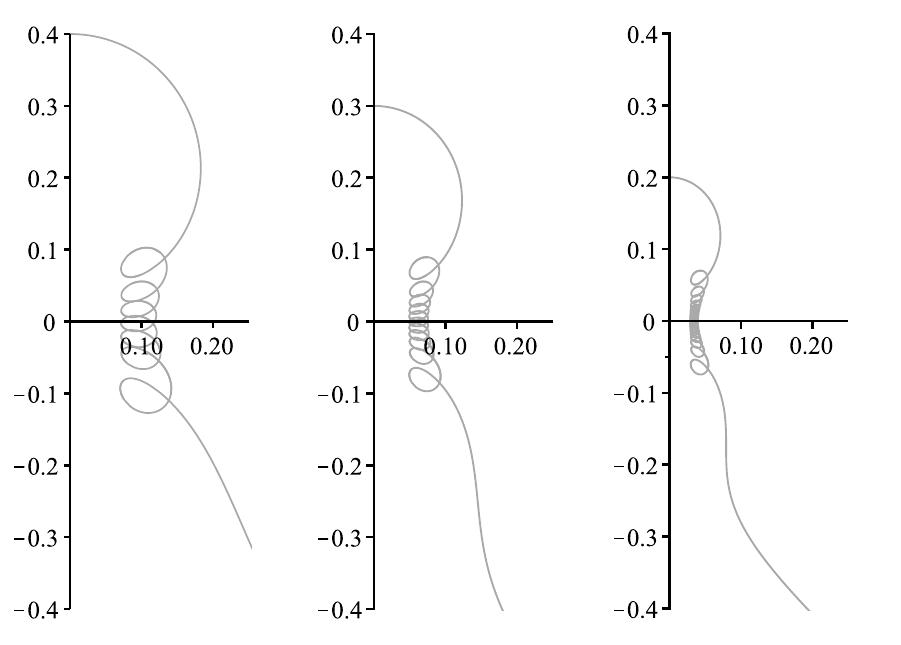}
\caption{Many-looped MOTOS. Notice that all three appear to exit towards infinity heading downwards. This is not coincidence (see Appendix \ref{Sec:AB}).  }
\label{LAL1}
\end{figure}

These steps appear to repeat ad infinitum
with shorter and shorter periods. Starting from a MOTS at $z_i$, for $z<z_i$ a loop forms and
 that loop migrates to join the other loops arrayed around $z=0$. While
this is happening the branch of the curve that heads out to infinity also pulls back towards the $z$-axis until it pinches off there to form a new MOTS at $z_{i+1}$
(with the other loops now symmetric around $z=0$). The first twelve 
 MOTS are shown in FIG.~\ref{FIG_MOTS12}. Some  higher loop MOTOS with intermediate $z_o$ values are shown in FIG.~\ref{LAL1}.

Thus as $z_o \rightarrow 0$ from above,  we expect more and more loops to be squeezed into smaller and smaller areas. However while curve complexity grows that complexity  is also more and more confined. We might then expect the limit curve to be relatively simple and indeed this seems to be the case. To leading order, note that
all of these curves are oriented upwards as they finish their loops and exit $r=2m$, and so asymptotically they must all approach $z=-2 \sqrt{2mR}$ in the same way that the $z_o < 0$ curves
did. In fact the similarity in the properties of the MOTOS as they
approach $z=0$ from above and below appear to be quite a bit stronger than just leading order asymptotics. They also appear to match at sub-leading order and are similar even for relatively small
values of $r$ (see Appendix \ref{Sec:AB}).

Hence, the divergence in behaviours between upward-oriented $z_o>0$ and $z_o< 0$ MOTOS may not be quite as different as it initially seemed. In fact there seems to be a continuity of many properties across $z_o=0$. 
%That continuity breaks down at $r=0$ itself (which is a discontinuity) and in that limit there is also a
%qualitative difference in behaviours between the infinite limit of loops above and absence of loops below.
% However  away from the singularity the curves are much more similar. 
 We will return to this point in Section \ref{merger}.

%
% Then on the other side there is a similar wrapping. However now the MOTOS turns by looping through a self-intersection before exiting $r=2m$. As $r_o$ continues to decrease that
% loop grows and moves towards $z=0$ and the part of the curve heading out to infinity in turn begins to pull back towards the $z$-axis. The turn from the $z$-axis 
% becoming sharper and sharper until for $r_o \approx 1.037m$ the curve intersects the $z$-axis at a right angle: this limit is a closed surface and so a (self-intersecting)  MOTS. It pairs with 
% the  upwards pointing surface  with $r_o \approx 1.037m$ from the last section to form the limit surface. 
% 
% On the other side of $r_o \approx 1.037m$, another loops forms which then moves progressively inwards towards $z_o$. At the same time, the part of the curve that escapes to infinity draws 
% in and develops a kink. From our numerical observations this process appears to continue {ad infinitum}: as $r_o  \rightarrow \infty$ more-and-more loops are added with the transition to each new 
% loop being a closed (self-intersecting) MOTS along with an upwards oriented curve that intersects the axis at $z<0$.

\subsubsection{Methods}

Before moving on to other results,  let us briefly examine the mechanics of the numerical integrations that provided these curves. Starting with the simplest case, for curves $z_o \gtrsim 2m$ 
there is a minimum value of $\theta$ (at the turn-around)
and so there is no single valued $R(\theta)$ that will fully cover it. Hence we necessarily switch from $R_{\mbox{\tiny{eq}}}^+$ to $\Theta_{\mbox{\tiny{eq}}}^+$ to find 
the full generating curve for the MOTOS. 

For the loops with $z_o \lesssim 2m_o$ curves, things are more complicated. 
Moving along the curve, starting from the positive $z$-axis, there are successive minima in $\theta$ and then $r$
 and we necessarily progress from $R_{\mbox{\tiny{eq}}}^+$ through 
 $\Theta_{\mbox{\tiny{eq}}}^-$ to $\Theta_{\mbox{\tiny{eq}}}^+$ to generate the full curve. These curves make full use of the allowed behaviours from Section \ref{Gen}. 

For multi-loop curves, there are many points where either $R_\theta$ or $\Theta_r$ vanish. Hence the integrations require repeated cycles through the 
$R_{\mbox{\tiny{eq}}}^+, \Theta_{\mbox{\tiny{eq}}}^-, R_{\mbox{\tiny{eq}}}^-, \Theta_{\mbox{\tiny{eq}}}^+$ family of equations (roughly one cycle for each loop) 
to generate the full curve.

\subsection{MOTOS that do not intersect the $z$-axis}
\label{nonZ}

Next we consider MOTOS that do not intersect the $z$-axis. This is necessarily a much broader class of surfaces than in the previous section (where there was just one parameter $z_o$ 
plus the orientation). However, as we saw  in Section \ref{LargeR}, their asymptotic behaviour is severely constrained to be parabolic (\ref{parabola}) and as we shall now see, their inner structures 
 turn out to be qualitatively similar to the closed MOTS of FIG.~\ref{FIG_MOTS12}.

\subsubsection{Perpendicular to the $x$-axis}

We start with surfaces that perpendicularly intersect the $x$-axis at some $x_o$. Then we are again looking at a one parameter family of oriented curves, this time with a reflectional symmetry through the $x$-axis. The techniques for finding these surfaces are the same as before: cycling through our four MOTOS equations in order to patch together a full picture of the surface. 

\begin{figure}
\includegraphics[scale=0.75]{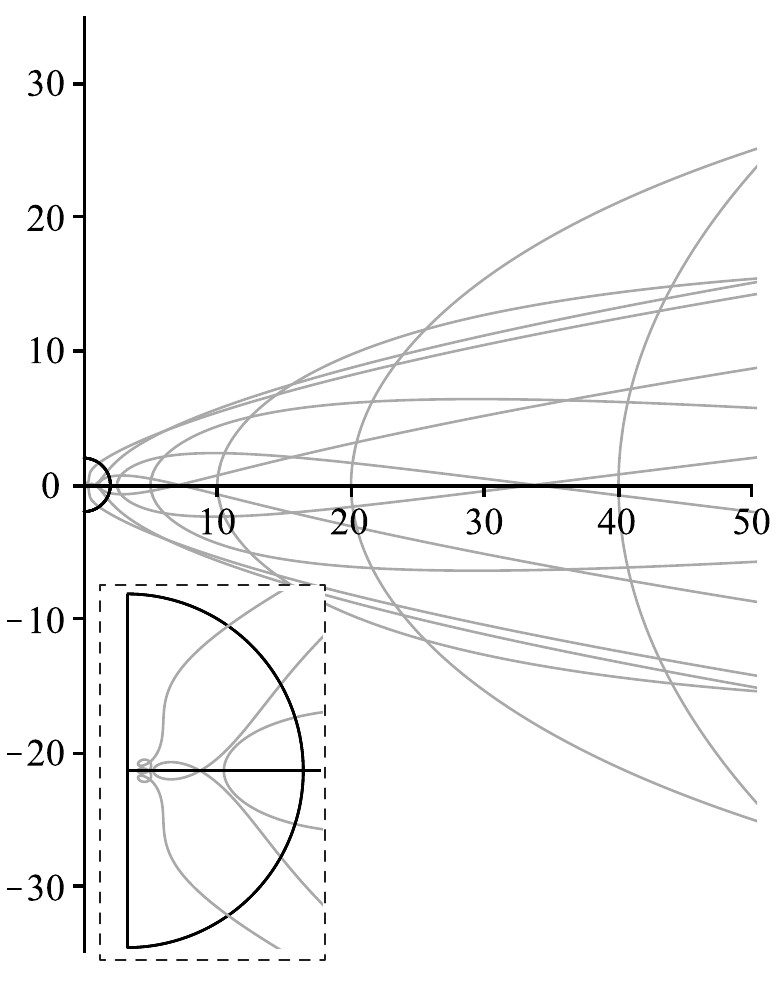}
\caption{Typical (initially)  inward oriented MOTOS that perpendicularly cross the $x$-axis. A blow-up of the region inside $r=2m$ is shown in the inset. }
\label{XIN}
\end{figure}

The behaviours of the MOTOS have familiar elements.  For $x_o > 2m$, the initially inward-oriented surfaces of FIG.~\ref{XIN} have the opposite orientation to $r=2m$ and so do not need to wrap around it
as $x_o \rightarrow 2m$. However the rules for large $r$ behaviour that we found in Section \ref{LargeR} still apply and tell us that the ``upward'' oriented branch 
asymptotes to $z = - 2\sqrt{2mr}$ while the ``downward'' oriented branch asymptotes to $z = 2 \sqrt{2mr}$. To achieve these asymptotics, the branches have to cross the $x$-axis at
some $x>x_o$ and this can be seen for some of the MOTOS in the figure (the ones with larger $x_o$ intersect beyond the range shown). These MOTOS are self-intersecting even while
being always outside $r=2m$. 

For $x_o < 2m$ the usual 
intricacies show up.  All the MOTS with an odd number of loops are also part of this family. This is not shown directly in the figure but is clear from an examination of FIG.~\ref{FIG_MOTS12}
where the MOTS with an odd number of loops intersect the $x$-axis with inward orientation. 

\begin{figure}
\includegraphics[scale=0.8]{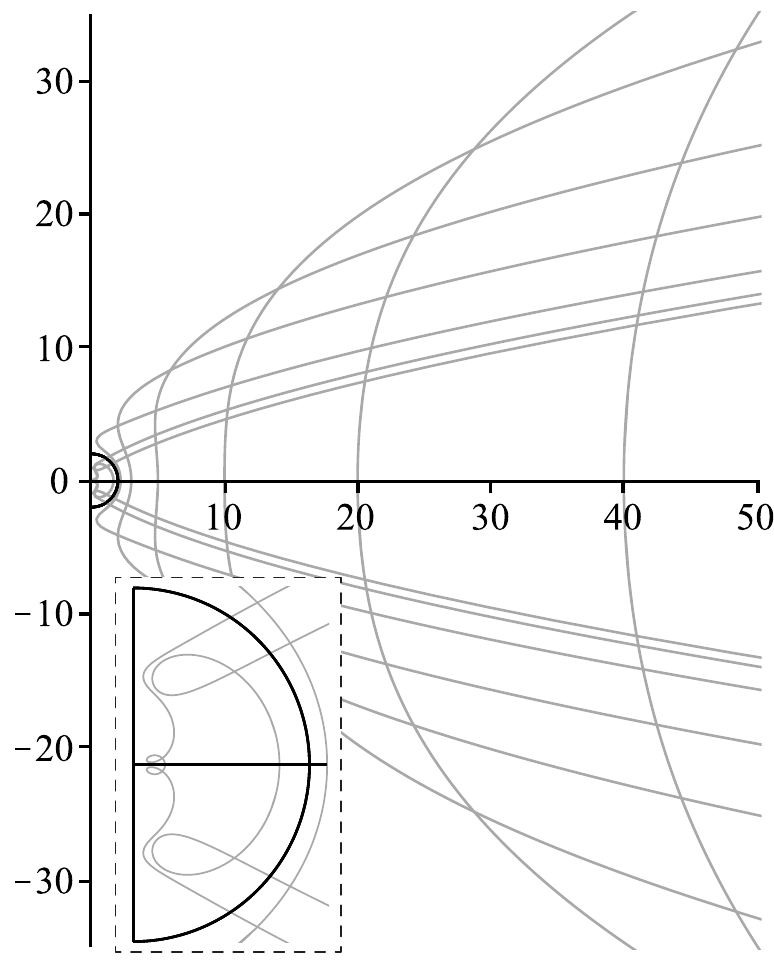}
\caption{Typical (initially) outward oriented MOTOS that perpendicularly cross the $x$-axis. A blow-up of the region inside $r=2m$ is shown in the inset. }
\label{XOUT}
\end{figure}

For the initially outward-oriented MOTOS shown in FIG.~\ref{XOUT} the original orientations of each branch match those required asymptotically so there is no need for  $x_o>2m$ surfaces to 
cross the $x$-axis and, in these examples, they do not. As $x_o \rightarrow 2m$ they wrap close to  $r=2m$ as we saw for those starting from the $z$-axis. Of course, $r=2m$ is part of this family. For 
$x_o < 2m$ we once again find looping surfaces. All MOTS with an even number of loops are also part of this family. Again they are not shown in the figure but this is clear from the orientation of 
the $x$-intercepts in FIG.~\ref{FIG_MOTS12}. 

%\begin{figure}
%\includegraphics{Fig_X_Out}
%\caption{}
%\label{XOut}
%\end{figure}

\subsubsection{Not-perpendicular to the $x$-axis}
So far we have considered one-parameter families of curves: those that intercept either the $x$-axis or $z$-axis at right angles. However there are also many curves that do not intercept
perpendicularly. These actually include all other possible (smooth) curves. A smooth curve that does not intersect the $z$-axis necessarily has two ends and asymptotically those ends will 
necessarily have opposite orientations: one up and one down. However by the considerations of Section \ref{LargeR}, asymptotically these must  end up on opposite sides of the 
$x$-axis. Hence  all of these curves have to cross that axis somewhere. 

This family includes all other curves but they are more difficult to study systematically  as  they are now parameterized by two numbers: the point and
angle of intersection with the $x$-axis. However from an initial study we do not think that these present any dramatically new behaviours. The
familiar elements are still all there. For those starting and remaining outside $r=2m$, the branches of inward-oriented ones are forced to self-intersect in order to have the correct asymptotic behaviour. Meanwhile, initially outward-oriented ones necessarily wrap close to $r=2m$ if they approach it. And, as we have come to expect, inside $r=2m$ the MOTOS can have multiple self-intersections. 
\begin{figure}
\includegraphics[scale=0.8]{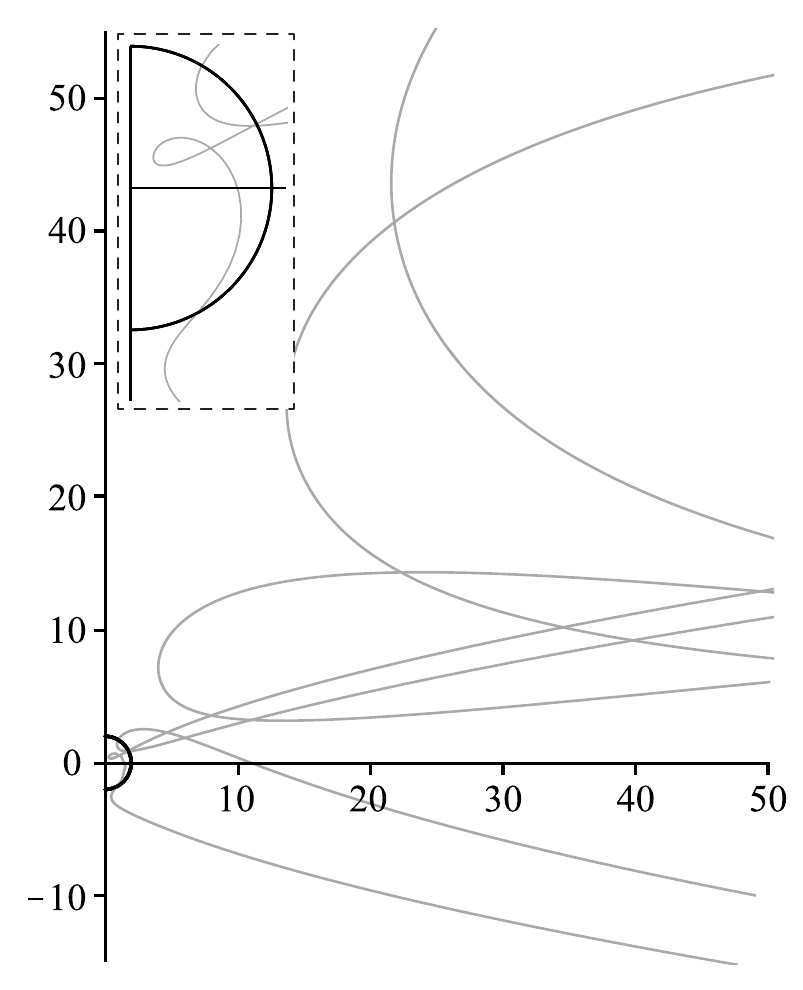}
\caption{Typical (initially) inward oriented MOTOS that are not symmetric around the $z$-axis (they have $R_\theta(\pi/4)=0$). A blow-up of the region inside $r=2m$ is shown in the inset. }
\label{X45in}
\end{figure}
\begin{figure}
\includegraphics[scale=0.8]{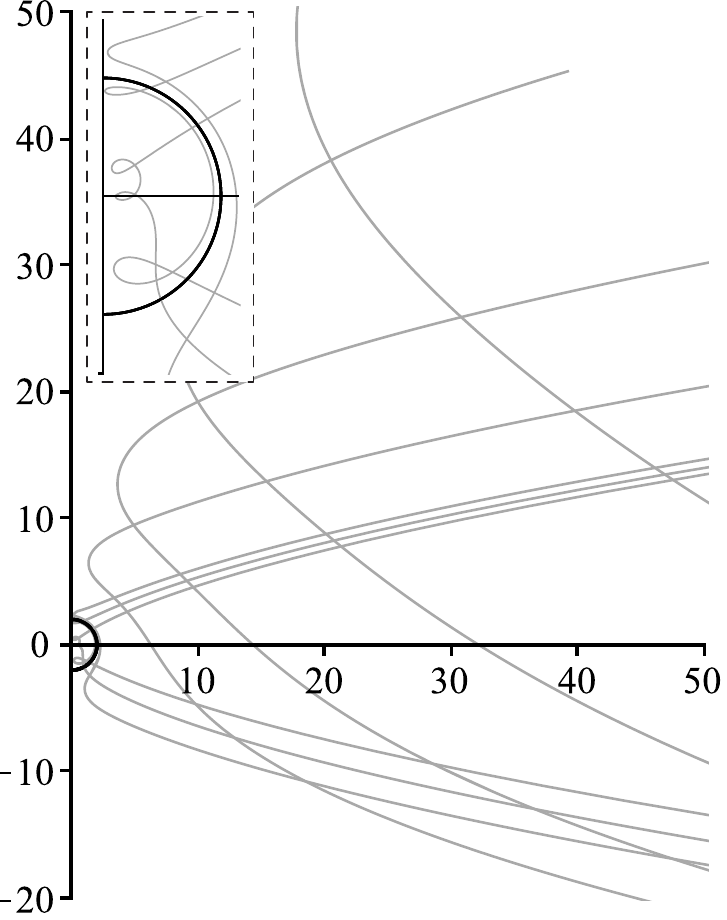}
\caption{Typical (initially) outward oriented MOTOS that are not symmetric around the $z$-axis (they have $R_\theta(\pi/4)=0$). A blow-up of the region inside $r=2m$ is shown in the inset. }
\label{X45out}
\end{figure}

%{\bf \dots \dots \dots}

\section{Extreme mass ratio mergers}
\label{merger}

As was discussed in the introduction, during a non-rotating extreme mass ratio merger, one would expect spacetime near the small hole to be very close to Schwarzschild. In that regime, full
spacetime MOTS should be (very close to being) sections of the MOTOS in Schwarzschild. Hence all stages of an extreme mass ratio merger in some neighbourhood of the small black hole should 
find representation among the families of surfaces studied in the last section. The difficulty is, of course, identifying which are the correct surfaces and then assembling them in the correct order. 

A time-ordered set of MOTS forms a marginally outer trapped tube (MOTT) and there are partial differential equations which define the evolution of such surfaces (see for example \cite{Andersson:2005gq,Andersson:2009aa,Jaramillo:2011zw}). 
Hence given a MOTS in one leaf, one can calculate its future evolution. However the equations are complicated (the time derivative for MOTS evolution is determined by the 
solution of an elliptic differential equation that must be solved on the MOTS) and while they extend to MOTOS, solving them is a non-trivial numerical problem and beyond the scope of this paper. 

We intend to return to such evolutions in future works. However even in their absence, the results of the last two sections still give rise to interesting results about the evolution of MOTS in 
extreme mass ratio mergers.
We have a range of possible behaviours, along with an understanding that some other behaviours are impossible; the possible evolutions are underconstrained but they are not unconstrained. 
In this section we will propose an evolution that is consistent both with the surfaces uncovered in the last sections as well as evolutions seen in full numerical studies such as \cite{Pook-Kolb:2018igu,Pook-Kolb:2019iao,Pook-Kolb:2019ssg,Evans:2020lbq}. 
However it should be kept in mind that what follows is not a rigorous evolution but is instead an informed speculation.
%however it should be emphasized that they are only informed guesses. 

%
% in the future works. 
%
%
%
%Even if this is a partial answer, it is not the full answer. In numerical simulations of  black hole mergers (see for example \cite{Anninos:1994ay,Bishop1982,Cadez1973,Cook:1992} or many recent 
%papers) one does not just see a simple evolution of a (pair of) MOTS throughout the merger: 
%instead at any point in the time there may be multiple MOTS and as time evolves they may appear and disappear. 
%
%Some of these appearances and disappearances are due to the vagaries of apparent horizon finders\footnote{See \cite{Pook-Kolb:2018igu,Pook-Kolb:2019iao,Pook-Kolb:2019ssg} for a good discussion of the difficulties of finding 
%complicated MOTS.}. 

%
%At this stage we don't have a good method for identifying preferred MOT(O)S. Nevertheless, based on the MOT(O)S that we have catalogued, we have a range of possible behaviours along
%with an understanding that some other behaviours are impossible. From the work of this paper, the possible evolutions are underconstrained but they are not unconstrained. 
%In this section we will propose a couple of possible evolutions that are consistent
%with the surfaces uncovered in the last sections. These proposals are particularly informed by the evolutions seen in \cite{Pook-Kolb:2018igu,Pook-Kolb:2019iao,Pook-Kolb:2019ssg}, however it should be emphasized that they are only
%informed guesses. 

\subsection{MOTS during the approach}
\begin{figure}[h]
\includegraphics[scale=1.0]{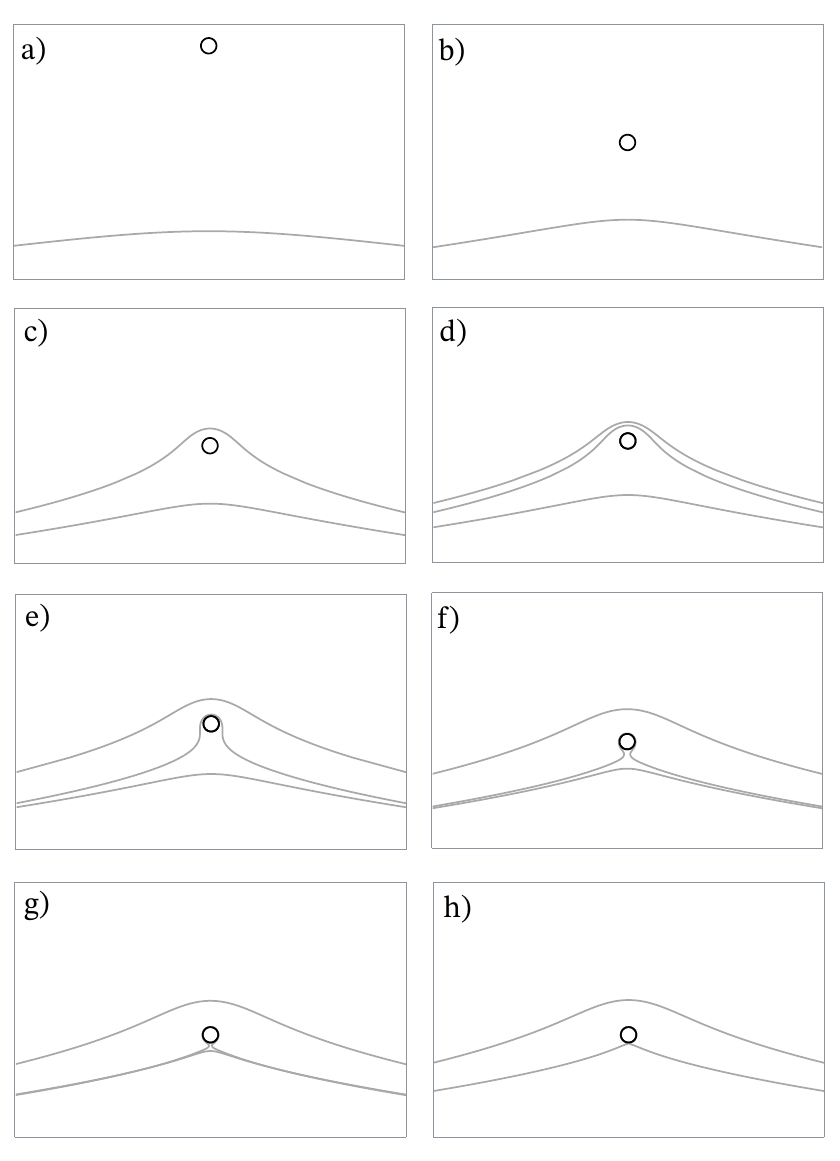}
\caption{Possible frames in a movie of the early stages of an extreme mass ratio merger. While all surfaces in these frames are (portions) of MOTOS from the earlier sections,
 they have been assembled by hand. Details should not be taken too seriously!}
\label{Infall1}
\end{figure}

Evolution up to the point where the large and small black hole MOTS touch has been studied  many times and is fairly well understood. 
FIG.~\ref{Infall1} shows a possible assembly of the pieces from the last few sections that is consistent with that understanding.  
Note that the $r=0$ centre of the coordinate system has been offset in successive images to make it clear 
that we are now thinking about a merger. 
%\footnote{Keep in mind though that all surfaces shown are just MOT(O)S in Schwarzschild and have been pieced together to present a possible merger 
%scenario. The techniques used in this paper can't provided true dynamics.}. 

Initially in a) the small MOTS is at $r=2m$ and the large black hole MOTOS is oriented upwards and towards it: the small
black hole is outside of it. Note that the large MOTS deforms up towards the small one and this deformation increases in b) as the 
small black hole gets closer. This is consistent with the behaviours observed in  \cite{Mosta:2015sga,Pook-Kolb:2019iao,Pook-Kolb:2019ssg}. 

There are also studies in which the large MOTS appears to deform away from the small black hole (\cite{Pook-Kolb:2018igu,Hussain:2017ihw} and the 
initial data surface shown in FIG.~2 of \cite{Mosta:2015sga}).
Such behaviour is impossible for us; all upward oriented MOTOS starting with $z_o<0$ deform towards the small black hole. 
However there is a significant difference between those papers which show 
deformations towards the small black hole versus those that show deformations away from it. Those deforming away study the MOTS in time-symmetric slices for which, as we saw earlier, the MOTS are minimal surfaces.
Hence the uniqueness theorem holds and one can think of the deforming away as being preliminary to the surfaces coinciding once they become tangent. By contrast, those papers in which 
the MOTS deforms towards the small black hole study these surfaces in non time-symmetric slices for which the degeneracy between inner and outer expansions is broken. 

%
%
%towards use non-time symmetric slices while those deforming away use time
%
%In both subfigures a) and b), note that the large black hole MOTOS puckers out towards the small one. While some
%previous papers have found that the large MOTS should initially pull back from the small black hole \cite{Pook-Kolb:2018igu,Hussain:2017ihw} 
%this is not a possibility for the surfaces that we are considering\footnote{}. 
%As seen in Section \ref{FromBelow}, all axisymmetric MOTOS oriented towards the small black hole bend towards it. The difference with the earlier mentioned papers is that they 
%worked in  coordinates for which MOT(O)S are minimal surfaces  and so have to ape $r=2m$ as they approach it. By contrast studies using horizon penetrating coordinates
%have seen the MOTS pucker out as we see here (for example \cite{Mosta:2015sga,Pook-Kolb:2019iao,Pook-Kolb:2019ssg}).
%
%See also the time-symmetric initial data surface shown in FIG.~2 of \cite{Mosta:2015sga}.

Going back to the figure,  c) proposes a jump of the large horizon MOTOS to encompass the small. In d) this bifurcates into an inner and an outer MOTS. Such jumps and bifurcations are 
commonly seen in numerical mergers and in particular this is consistent with \cite{Mosta:2015sga,Pook-Kolb:2019iao,Pook-Kolb:2019ssg}. 

The choice of the jump surface was not arbitrary. Horizon jumps are generally identified with locations where a MOTT is instantaneously tangent to the time foliation 
\cite{BenDov:2004gh,Booth:2005ng, Ziprick:2008cy,Jaramillo:2009zz,Chu:2010yu,Bousso:2015qqa,Gupta:2018znn,Pook-Kolb:2019iao,Pook-Kolb:2019ssg}. Such a MOTS is also 
instantaneously ``extremal'' in the sense that small deformations  both inwards and outwards can transform it into an untrapped surface. For MOTOS we propose that the
equivalent condition be that such a deformation exist and the magnitude of the generating vector be bounded: most MOTOS will not satisfy this requirement as they and their neighbouring 
MOTOS will  diverge as $r \rightarrow \infty$. However the MOTOS originating from $z_o \approx 4.45m$ does meet this requirement. In Appendix \ref{Sec:AB} it is shown  to be at a turning point in 
the asymptotic behaviour, which will be sufficient to ensure that the magnitude is bounded.

%The choice of the ``jump'' surface in c) was not arbitrary. As discussed in Appendix \ref{Sec:AB},
%this is a surface for which the first variation of $\theta_+$ in the $\hat{r}$ direction vanishes. This is analogous to the condition that such surfaces necessarily satisfy when the MOTT becomes tangent  to the leaf (again see FIG.~2 of \cite{Gupta:2018znn}). 

Returning to the figure, e) to g)  show the inner MOTOS starting to wrap around the small black hole (as it must since they are oriented in the same direction on the $z$-axis) while the outer MOTOS relaxes. The wrapping becomes tighter and tighter until in h) it coincides with the original large black hole MOTOS (which now has a cusp) plus $r=2m$. This is analogous to the behaviour that is seen in, for example,
\cite{Jaramillo:2009zz,Pook-Kolb:2019iao,Pook-Kolb:2019ssg}.

\subsection{MOTS during the late merger}
\begin{figure}
\includegraphics[scale=0.95]{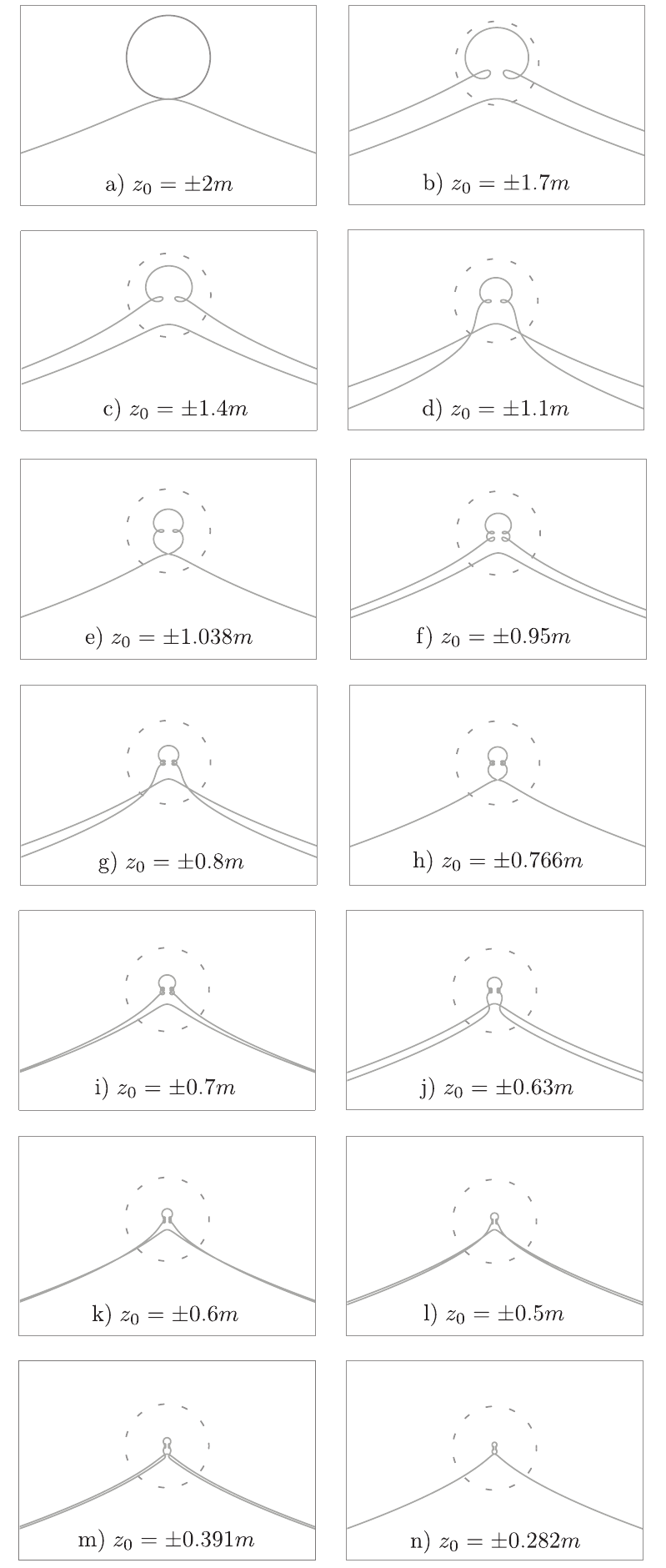}
\caption{Possible frames in a movie of the later stages of an extreme mass ratio merger, zooming in on the small black hole. 
For post-a) subfigures the $r=2m$ MOTS is drawn as a dashed line however
the other, self-intersecting, MOTS are only shown when they appear as part of the sequence.  }
\label{Infall2}
\end{figure}
As noted, the evolution shown in FIG.~\ref{Infall1} is not controversial. Most black hole merger calculations show such jumps and horizon creations. However moving beyond the point of tangency we enter lesser known territory. FIG.~\ref{Infall2} picks up at that point and focuses on the region close to $r=2m$. The outer MOTOS is not shown in this figure but we expect that it would continue to relax. 

This second proposed set of frames is organized around three observations. 
\begin{enumerate}
\item From Section \ref{nonZ}, any MOTOS that does not intersect the $z$-axis necessarily has one  (upward-oriented) end asymptotic to $z=-2 \sqrt{2mr}$ and the other 
(downward-oriented) end asymptotic to $z=2 \sqrt{2mr}$.
\item As we saw in Section \ref{FromAbove}, if $z_i>0$ generates a self-intersecting MOTS (as in FIG.~\ref{FIG_MOTS12}) then it, along with  the tangent 
upward-oriented MOTOS starting from $-z_i$, combine to form the limit curve for upward oriented  $z_o \rightarrow z_i$ MOTOS (from above and below). 
\item All upward-oriented MOTOS have the same $z=-2 \sqrt{2mr}$ asymptotic behaviour. However in Appendix \ref{Sec:AB} it is argued that as $z_o \rightarrow 0$ from 
above, the sub-leading order terms may also match. Asymptotically the $z_o>0$ and $z_o<0$ MOTOS approach the same curve. 
Further as discussed in Section \ref{FromAbove}, the $z_o>0$ complexities are confined
in smaller and smaller regions as that happens. 
\end{enumerate}

We interpret the first observation as indicating that only MOTOS that intersect the $z$-axis should be considered in modelling extreme mass ratio mergers. Including a two-ended MOTOS
would mean a MOTOS running to infinity as $z=2 \sqrt{2mr}$ and being oriented downwards. This is not consistent with the physical situation: a MOTOS should be associated 
with either the original large black hole or the merged black hole, but such a surface would be oriented in the opposite direction from the original large black hole MOTS. 

Next, the second observation means that  a subset of the $z_o<0$ upward-oriented MOTOS are (in the limit) part of the set of  $z_0>0$ upward-oriented MOTOS. The cleanest way that this 
can happen is if the $z_o<0$ MOTOS arrived at each $-z_i$ at exactly the same time as the $z_o>0$ MOTOS arrive at $z_i$. As a working hypothesis we will assume that this is 
true. Among other nice properties it provides us with milestones at which we can match the $z_o>0$ and $z_o<0$ evolutions, even if we are not sure how the evolution happens between those
milestones. 

The third observation is essentially the $z_o \rightarrow 0$ limit of the second. By this observation both above and below MOTOS should arrive at $z_o = 0$ at the same time and (possibly)
annihilate at that point leaving only $r=2m$ (and presumably the other closed MOTS)  inside the outer horizon. 

% as $z_o$ decreases we automatically work 
%through the family of MOTS.  Then, continuity suggests that the upward oriented MOTOS should arrive at the same time so as to provide the full limit curve. 
%This then allows us to coordinate the approach of the inner and original large black hole MOTOS to the singularity. Such meetings (or near meetings) are seen in FIG.~\ref{Infall2} a), e), g) and m).  
%
%If we accept this sequence there is a strong constraint on how the MOTOS can evolve. However between these fixed points the matching is not so obvious. For now 
%we have simply chosen to plot the upward oriented $\pm z_o$ together. 
From these starting points we arrive at the evolution shown in FIG.~\ref{Infall2}, where for simplicity we have extended the $\pm z_i$ milestone matching to all values. The
true relationship may be more complicated, but if the milestones are correct it cannot be too different, especially for the smaller values of $z_o$ where the  $z_i$ become closer and 
closer together.

Regardless of these details, the big picture is that as the inner MOTS contracts with $z_o \rightarrow 0$ from above, it develops more and more loops and those loops become both tighter
and closer together. In the limit approaching the singularity, there are an infinite number of loops packed infinitely close together.  At the same time, away from the singularity, the MOTOS 
approaches the same limit curve as when $z_o \rightarrow 0$ from below for the original large black hole MOTS. 
We propose that when they meet they annihilate and we are left with the small black
hole. This would then continue to move into the interior of the large black hole while the outer horizon (not shown in these diagrams) relaxes.

This sequence of events is consistent with the full numerical simulations of  \cite{Mosta:2015sga} and \cite{Pook-Kolb:2019iao,Pook-Kolb:2019ssg}. In \cite{Mosta:2015sga} the large
black hole MOTS  can be seen developing a sharp point  as the small black hole singularity approaches (their FIG.~5 and FIG.~7). Unfortunately that is as far as that simulation could track the MOTS
and so our proposed later steps cannot be compared. A longer comparison can be made with \cite{Pook-Kolb:2019iao,Pook-Kolb:2019ssg}. 
In those papers comparable-mass black holes are considered. Focusing on the more recent \cite{Pook-Kolb:2019ssg}, their FIG.~1 shows the jump to pair create inner and outer MOTS, followed by 
the contraction of the inner horizon around the original MOTS and then the formation of a self-intersecting MOTS (the equivalent of the loops in our FIG.~\ref{Infall2} b{\small)}). This is as far as 
the horizon finder of that simulation could follow the MOTS.

\section{Discussion}
\label{discuss}

%We have seen that the possible MOTOS in Painlev\'e-Gullstrand slices are varied but  constrained. Among other things their free ends have (leading order) universal asymptotics 
%that depend only on the orientation, they can only intersect the $z$-axis at a right angle and maxima in $r$ can only occur inside $r=2m$. These constraints mean that outside of $r=2m$, MOTOS
%can only have fairly simple behaviours and any exotica are constrained to remain inside $r=2m$. That said we have also seen that in that region, the MOT(O)S take full advantage of their
%potential freedoms and can develop an arbitrary number of self-intersections. 

The observations of the preceding sections raise at least as many questions as they answer. Here we discuss a few of these issues. 

\subsection{Self-intersecting MOTS}

We have found  an (apparently) infinite number of self-intersecting MOTS inside $r=2m$.
As far as we are aware, 
such interior MOTS have not been previously observed in the Schwarzschild spacetime. While a full study of the geometry of these surfaces will be left for an upcoming paper, here we 
note that the MOTT generated by taking a particular $n$-loop MOTS in each surface of constant $\tau$ is neither an isolated nor a dynamical horizon\cite{Ashtekar:2004cn}. 

To see this, first consider the three-surface $(\tau, R(\lambda) , \Theta(\lambda) , \phi)$ that is generated by propagating a 
general axisymmetric two-surface $(R(\lambda) , \Theta(\lambda) , \phi)$ onto all surfaces of constant $\tau$. 
Then that surface has induced metric:
\begin{align}
 q_{ij} \dd y^i \dd y^j  = &  - \left( 1- \frac{2m}{R} \right) \dd \tau^2 + 2 \sqrt{\frac{2m}{R}} \dot{R} \dd \tau \dd \lambda \\
& + (\dot{R}^2 + R^2 \dot{\Theta}^2 ) \dd \lambda^2 + R^2 \sin^2\!  \Theta \dd \phi^2 \nonumber
\end{align}
which has determinant
\begin{align}
\mbox{det} (q_{ij}) =  \left( R(2m-R) \dot{\Theta}^2 - \dot{R}^2 \right) R^2 \sin^2  \!\Theta \; .  \label{det}
\end{align}
This is positive, zero or negative at locations where the surface is spacelike, null or timelike respectively. 

The first thing to note is that, besides the horizon, there are no consistent solutions that are purely null. From (\ref{det}) the surface is everywhere null only if either $R=2m$ or 
\be 
R(\theta) = \frac{2 m \tan^2 \left[(-\theta + C)/2 \right]}{1+\tan^2 \left[(-\theta + C)/2 \right]} \, . 
\ee
This second possibility can easily seen to be inconsistent with~\eqref{REq} unless $m=0$. Hence $R=2m$ is the only such null MOTT. 

Returning to (\ref{det}) we note that for $R>2m$, any such constant geometry, axisymmetric surface is necessarily timelike. However for $R<2m$ the signature can be locally spacelike, timelike or null and can change
as a function of $\lambda$.
In particular if $R<2m$, then wherever $\dot{R}=0$ the surface is spacelike and wherever $\dot{\Theta}=0$ it is timelike. 
Our self-intersecting MOTS all have maxima and minima of both $R$ and $\Theta$ and so the corresponding MOTT for each of these surfaces has timelike, null  and spacelike sections. 
Hence the MOTTs are neither isolated nor dynamical horizons. 

The full geometry of the MOTS (including their stability) and their associated MOTTs will be studied in a future paper. 

\subsection{Robustness of observations}
In this paper we have presented results for axisymmetric MOT(O)S in a single time foliation for a single spacetime. Hence the generality of the results is not clear. 
There are many obvious questions. 
Do all black holes harbour infinite numbers of MOTS in their interiors? If so do they only exist in special time foliations? If not do they exist for all stationary black holes? Is axisymmetry required? Is asymptotic structure significant? How would things change for a black hole with both outer and inner horizons (like Reissner-Nordstr\"om)?

We expect the existence of interior self-intersecting MOTS to be quite general and that they will exist in axisymmetric spacetimes %regardless of 
whether or not they are dynamic and independent of 
the asymptotic structure. These results should not be critically dependent on Schwarzschild spacetime. 
This proposal is supported by the simulations of 
\cite{Pook-Kolb:2019iao,Pook-Kolb:2019ssg} which first identified self-intersecting MOTS during a head-on merger. Also in \cite{Booth:2017fob} the qualitative behaviours of the MOTOS 
in the proximity of the standard outer horizon MOTS appeared to be the same for the whole Reissner-Nordstr\"om-deSitter family.\footnote{But keep in mind that in that reference 
as one moves deeper into the interior after the first turn from the $z$-axis, the remaining parts of the MOTOS 
are incorrect.} We also do not expect the asymptotic structure to play much of a role (since
these surfaces are likely to remain enclosed in any black hole). Whether axisymmetry is critical is unclear. The initial conditions for the surfaces certainly require fine tuning in order to close
and it seems possible that the loss of symmetry might disrupt this. On the other hand there are many more non-axisymmetric surfaces so the increase in freedom may balance off the 
fine tuning problem.

We expect that there will be self-intersecting MOTS in other horizon-penetrating coordinate systems and we also 
expect them to be constrained inside $r=2m$ (for Schwarzschild). For non-penetrating time foliations the situation 
is of course quite different. By their nature one cannot find horizon-crossing MOTOS in those slices but also the $r>2m$ MOTOS may be quite different. 
%An example of such a foliation is standard Schwarzschild time and restricting our attention to $r>2m$ the MOTOS are quite different from those we have seen so far. 
While preparing the current paper, we also studied the MOTOS in the usual time-symmetric Schwarzschild foliation and so can briefly present
an example of a class of  dramatically different MOTOS.

Recall from Section \ref{general} that in Schwarzschild time slices the MOTOS are minimal surfaces. This eliminates the distinction between inward and outward orientations and in particular
means that no two MOTOS can be tangent without being identical. Now focus on region of spacetime close to $r=2m$ using the coordinates
\begin{eqnarray}
r & = & 2m (1+ e^\rho) \; \; \mbox{and} \\
\theta & = & \frac{\pi}{2} + \arctan ( \sinh \xi )  \; . \nonumber
\end{eqnarray}
These coordinates blow-up the region spacetime close to the horizon ($r=2m$ is at $\rho = - \infty$) and are adapted for curves which may make sharp turns very close to the 
$z$-axis ($\lim_{\theta \rightarrow 0} \xi = -\infty$ and $\lim_{\theta \rightarrow \pi} \xi = \infty$). Though we do not provide details here, solving the minimal surface equations proceeds
by essentially the same methods that we used to find  MOTOS in this paper,  though they are a little less complicated as there is no need to worry about orientation.
Hence there are only two equations
to cycle through. 

\begin{figure}
\includegraphics[scale=0.45]{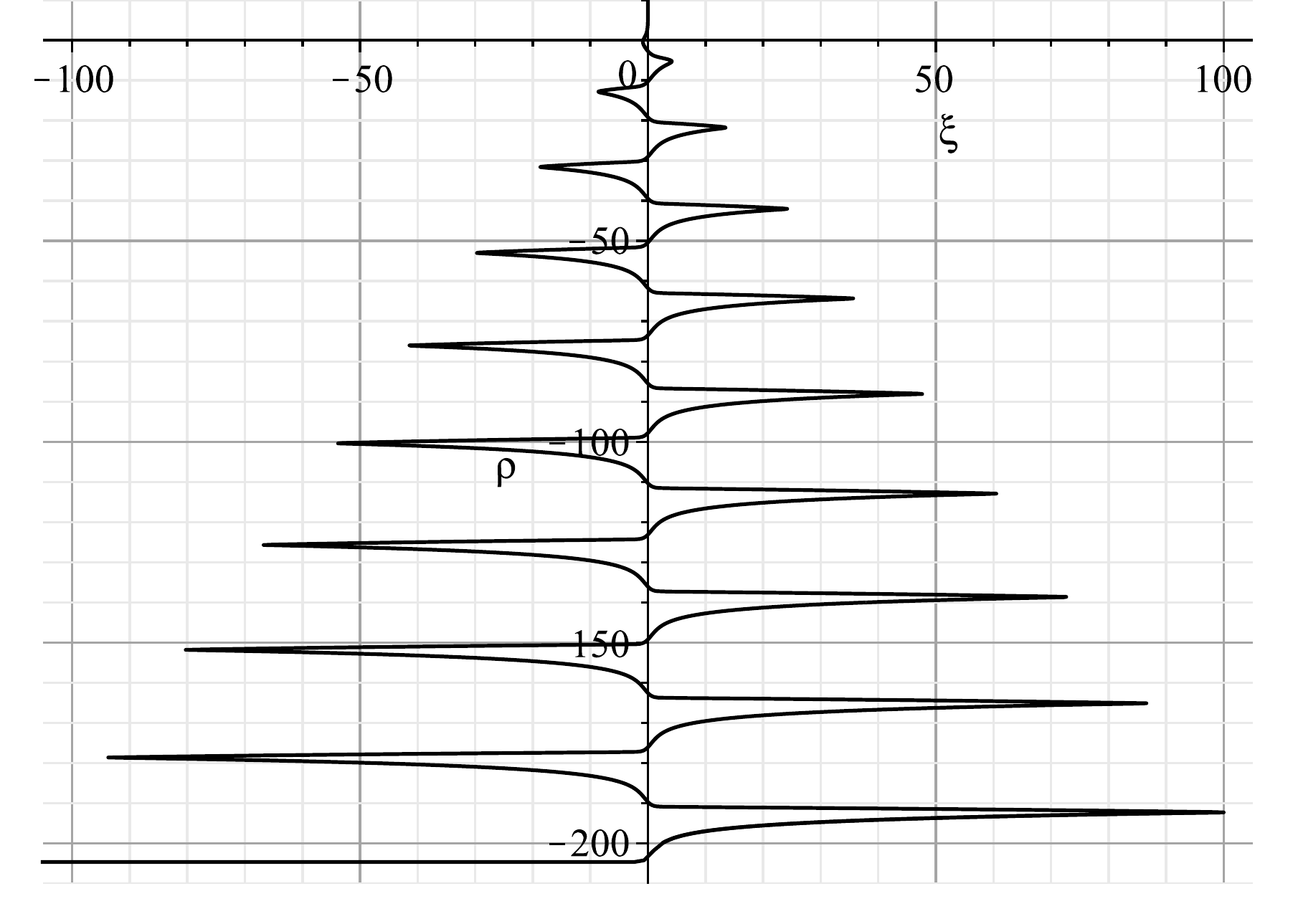}
\caption{Multiple wrappings of a minimal surface close to $r=2m$ in the Schwarzschild foliation. To get a feeling for the scales note that for 
$\xi = 20$, $40$, $60$, $80$ and $100$,
 $\theta \approx \pi$ to $8$, $17$, $25$, 
$35$ and $43$ decimal places respectively. Meanwhile for $\rho = -10$, $-20$, $-40$, $-80$ and $-160$, $r \approx 2m$ to $4$, $8$, $17$, $34$ and $69$ decimal places respectively. 
}
\label{CloseIn}
\end{figure}

As in the current study, surfaces departing from the $z$-axis must do so at a right angle. If they start close to $r=2m$ they only  slowly depart from it (like consistently oriented surfaces
in this paper). Similarly they must turn around before reaching the $z$-axis but then something quite different happens. After the turn they are still consistently oriented
with $r=2m$ (since they are minimal surfaces) and so continue to only slowly retreat. If they are sufficiently close this process can repeat an arbitrary number of times and so generate a MOTOS with an 
arbitrary number of folds wrapped close to $r=2m$. Such a multifold surface is shown in FIG.~\ref{CloseIn}. Equivalent surfaces have been found in Schwarzschild-AdS \cite{Hubeny:2013gta}.
 
It is certainly possible that other foliations might harbour other exotic MOT(O)S. 

% {\bf how robust are these findings?}

%{\bf not isolated - properly characterize}

\subsection{Extreme mass ratio mergers}

While the application to extreme mass ratio mergers is suggestive, many issues remain to be addressed. 

As noted, it should be possible to dynamically evolve MOTOS but doing that is highly non-trivial. In this paper we have instead
attempted to assemble the possible MOTOS into an evolution based on physical considerations and full numerical studies. This 
 presents a possible time evolution but checking whether or not it is correct will require significantly more work. 
 
 Apart from evolving a given MOTOS in time there is also the problem of identifying which are the initial surfaces from which we should evolve. A possible filter for 
 identifying  the ``correct'' MOTOS is that they should asymptote to the event horizon~\cite{Emparan:2016}. Here the idea is that as one gets far from the influence of 
 the small black hole, the event and apparent horizons should coincide as they do for stationary spacetimes. Our initial calculations show that the event horizon
 does indeed have the same leading order asymptotic behaviour as the upward-oriented MOTOS. So it is possible that this may work with the sub-leading order 
 terms selecting a MOTOS. However keep in mind that for much of the evolution there will be multiple MOTOS to identify, so this is likely not the full solution. 
 %Further, given that both event horizons and MOTS require global knowledge to be identified and hence knowledge of spacetime beyond the regime of the 
 %Schwarzschild approximation, it is not certain that this is the correct solution. 

% We have made an
% argument for where a horizon pair-creation event should appear but it admittedly ad hoc. In general a possible filter for identifying the ``correct'' MOTOS
%might be that they should asymptote to an event horizon calculated in the same way as in \cite{Emparan:2016}. Our calculations on this will be 
%presented elsewhere  

For this and other reasons, the asymptotic properties of the MOTOS still need to be better understood. While it is reasonable to argue
that close to the small black hole we should be able to understand MOTS as Schwarzschild MOTOS, it is not so clear what one should
do for small $z$ but large $x$. In our approximation the large black hole is represented by a MOTOS, but in the full solution it closes
up far from the small MOTS. In that asymptotic regime there are competing limits and 
for large $x$ one might  expect corrections to the 
Schwarzschild approximation. This correction is ignored in the main part of this paper but hinted at in the complicated asymptotics of Appendix \ref{Sec:AB}. 
For example, the MOTOS in each frame of FIG.~\ref{Infall2} share leading order asymptotics but often intersect outside of the frames. 

\acknowledgements

We would like to thank Jose Luis Jaramillo, Badri Krishnan, Daniel Pook-Kolb and Roberto Emparan for discussions, comments and suggestions. IB and SM were supported by the Natural Science and Engineering Research Council of Canada Discovery Grant 2018-0473. The work of RAH was supported by the Natural Science and Engineering Research Council of Canada through the Banting Postdoctoral Fellowship program.

%Finally 
%
%{\bf which ones should be included anyway?}

%
%
%{\bf area increase?}
%
\appendix

\section{Second order asymptotic behaviour for MOTOS intersecting the $z$-axis}
\label{Sec:AB}
In this appendix we consider the  large-$r$ behaviours of the MOTOS. From  (\ref{AsymptZ}) we know that 
the leading order asymptotic structure is universal with $z \approx \pm 2 \sqrt{2mr}$. However, we can extract the second-order (constant and $\log r$) corrections by 
subtracting that leading order term and plotting what remains as a log-linear graph. 

\begin{figure}
\includegraphics{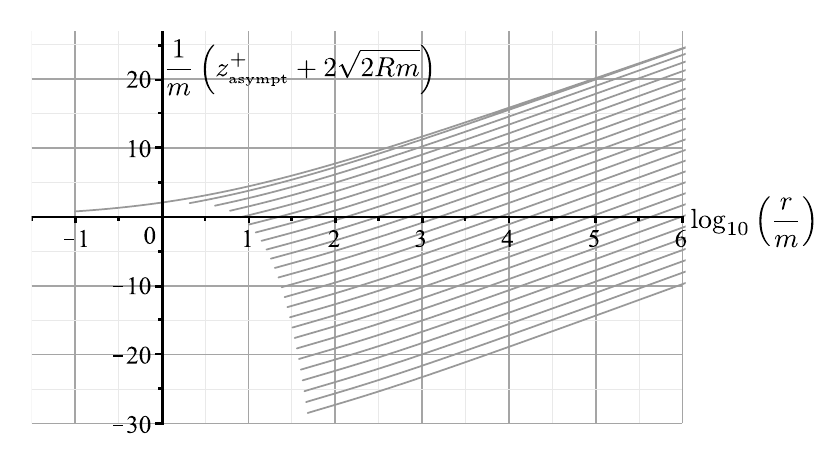}
\caption{Asymptotic behaviour of upward oriented rotationally symmetric MOTOS that intersect the $z$-axis with $z_o<0$. The vertical axis is the $z^+_{\mbox{\tiny{asympt}}}$ coordinate with 
the universal $-2\sqrt{2mR}$ asymptotics removed. The starting location for the surfaces ranges from $z_o =-0.1m$ to $z_o = -48.1m$ (hence their 
staggered starting points). As expected, their asymptotic behaviour is log-linear.  }
\label{FBA}
\end{figure}

First consider the upward-oriented MOTOS with $z_o<0$ from Section \ref{FromBelow} which are plotted in FIG.~\ref{FBA}.  As might be expected from FIG.~\ref{BelowLarge}, these MOTOS foliate their region of 
spacetime, arriving at infinity in the same ordering with which they left the $z$-axis. With the universal, leading order $2 \sqrt{2mR}$ removed the curves are asymptotically log-linear in  
$(z_{\mbox{\tiny{asympt}}}^+ + 2 \sqrt{2mr})$ plotted as a function of $r$. Further they appear to be parallel (up to $r = 10^6 m)$. From the figure, surfaces which start out 
below $- 2 \sqrt{2mr}$ will all cross it for sufficiently large $r$. 
%\begin{figure}
%\includegraphics{StraightLines}
%\caption{Asymptotic behaviour of upward oriented rotationally symmetric MOTOS that intersect the $z$-axis with $z<0$. The vertical axis is the $z^+_{\mbox{\tiny{asympt}}}$ coordinate with 
%the universal $-2\sqrt{2mR}$ asymptotics removed. The starting location for the surfaces ranges from $r_o =0.1m$ (to curve) to $r_o = 48.1m$ (bottom curve) with a spacing of
%$\Delta r_o = 2$. As expected their asymptotic behaviour is log-linear.}
%\label{Bel}
%\end{figure}

\begin{figure*}
\includegraphics{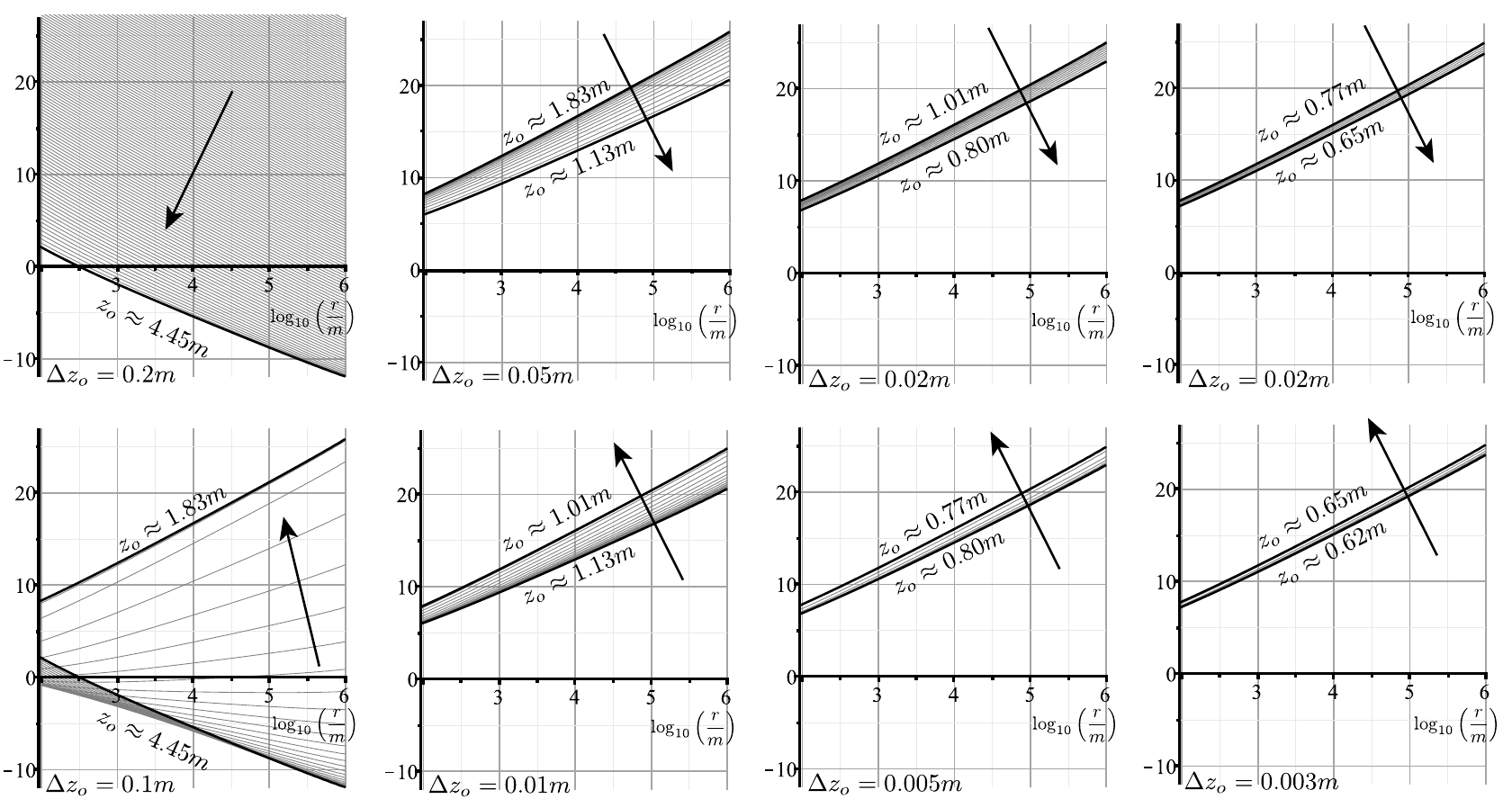}
\caption{Asymptotic behaviour of upward oriented MOTS starting from $z_o>0$. As in FIG.~\ref{FBA}, the vertical axis measures $z^+_{\mbox{\tiny{asympt}}} = R \cos \theta + 2 \sqrt{2mR}$ and as usual 
units are $m$. Arrows indicate the direction of decreasing $z_o$ (that is approaching $r=0$). Note that the progression of figures with $z_o$ decreasing is 
up-down, up-down, up-down, up-down. The spacing of surfaces is noted in the bottom left corner of each sub-figure.  %Numerical values of turning points are approximate.   
}
\label{FAA}
\end{figure*}

For (initially) upward oriented MOTOS with $z_o>0$ the situation is more complicated. As was seen in FIG.~\ref{FAL}--\ref{Evolve2}  the MOTOS can cross. Asymptotic behaviours are shown in FIG.~\ref{FAA}. This is a somewhat complicated figure and so needs to be interpreted with 
some care. 

%\begin{figure}
%\includegraphics[scale=0.7]{Fig_AsymptBelow}
%\caption{Asymptotic behaviour of upward oriented MOTS starting $z_o<0$. As in FIG.~\ref{FBA} The vertical axis is the $z^+_{\mbox{\tiny{asympt}}} = R \cos \theta + 2 \sqrt{2mR}$ and as usual 
%units are $m$. The arrow indicates the direction of decreasing $z_o$ (that is towards $-\infty$). The topmost surface is $z_o=-0.1m$ and then each subsequent surface decreases by $2m$.  %Numerical values of turning points are approximate.   
%}
%\label{FAB}
%\end{figure}

As in FIG.~\ref{FBA}, this figure plots the corrections to the leading order $z^+ \approx -2\sqrt{2mr}$. The first thing to notice is that for $z_o \gtrsim 4.45m$ the $z$-coordinate
of the MOTOS is decreasing relative to $-2\sqrt{2mr}$ while for $0<z_o \lesssim 1.83m$ it is increasing (as was the case for $z_o<0$). Further in these regimes the log-linear 
relationship appears to hold. While remaining positive, the slope of the curves oscillates, periodically becoming more or less steep but appearing (from these numerical
observations) to do so in a tighter and tighter range. 

For $1.83m \lesssim z_o \lesssim 4.45m$ the situation is less straightforward. Here the log-linear relationship breaks down. However this is not surprising as it is also in this regime 
where the asymptotic behaviour switches from decreasing to increasing (relative to $-2 \sqrt{2mr}$). During that transition, $\beta$ in (\ref{beta}) will necessarily transition through zero
and so the lower order (non log-linear) terms will temporarily become dominant. 

%{\bf (4.45 as the transition point?)}

What is the reason for the slope oscillations? They appear to be a by-product of the development of the loops. Comparing FIG.~\ref{FIG_MOTS12} and FIG.~\ref{FAA} it can be 
seen that the switch to decreasing slopes (top row of FIG.~\ref{FAA}) closely follows the formation of each new MOTS. From the first two such formations shown 
FIG.~\ref{FAL} and FIG.~\ref{Evolve1}, that decrease happens as the new loop forms which tilts slightly downwards compared to before the formation. However as the loop subsequently moves
away from its site of formation its tail  tilts up until the next loop forms when it tilts down again. Qualitatively this is the basic mechanism. 

There is another important feature of FIG.~\ref{FAA}. Although the asymptotics oscillate, the range of those oscillations decreases as the number of loops increases. Each range
is fully contained in that of the previous oscillation and the upper and lower bounds appear to be converging. Significantly, all of the ranges that we have checked contain the $z_o \rightarrow 0$ 
limit curve from FIG.~\ref{FBA} and for the larger values it appears to approach the upper bound of those ranges. 

\begin{figure}
\includegraphics{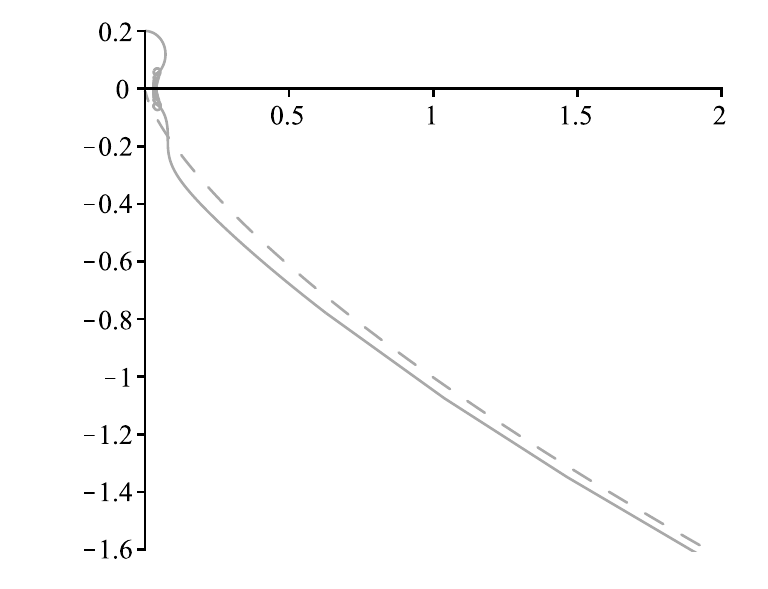}
\caption{The second curve from FIG.~\ref{LAL1} plotted next to the (dashed) upward oriented curve originating from $z_o=-0.01$.}
\label{AAB}
\end{figure}

Hence it appears that there is a continuity in the asymptotic behaviours of the upward oriented curves as $z_o \rightarrow 0$ from above and below: both have the same limiting curve. In fact
if one plots a few sample curves that limit seems to also hold for smaller values of $r$ (see for example FIG.~\ref{AAB}). That said, the limit can't be smooth over the entire curve: $r=0$ is 
a singularity and as $z_o \rightarrow 0$ from above we expect an infinite number of loops. 

\bibliographystyle{apsrev4-1} 
\bibliography{horizon}

\end{document}